%% file: paper.tex
\documentclass[letterpaper,twocolumn,10pt]{article}

\usepackage{usenix-2020-09}
\usepackage[english]{babel}
\usepackage{blindtext}
\usepackage{def}
\usepackage{package}
\usepackage[small,compact]{titlesec}

\begin{document}
\frenchspacing
\title{\Large \bf Unlocking the Power of Inline Floating-Point\\
  Operations on Programmable Switches}

\author{
{\rm Yifan Yuan}\\
UIUC
\and 
{\rm Omar Alama}\\
KAUST 
\and
{\rm Jiawei Fei}\\
KAUST 
\and  
{\rm Jacob Nelson}\\
Microsoft Research 
\and 
{\rm Dan R. K. Ports}\\
Microsoft Research 
\and 
{\rm Amedeo Sapio}\\
Intel 
\and 
{\rm Marco Canini}\\
KAUST   
\and 
{\rm Nam Sung Kim}\\
UIUC
} 

\date{}

\maketitle
\input{abstract}
\input{introduction}

\input{background}

\input{design}
\input{arch}
\input{ml}
\input{db}
\input{related}
\input{conclusion}

\clearpage
\bibliographystyle{abbrv}
\bibliography{references}

\clearpage
\input{appendix}

\end{document}

%% file: abstract.tex
\begin{abstract}

\footnote{This work has been accepted by a conference. The authoritative version of this work will appear in the Proceedings of the 19th USENIX Symposium on Networked Systems Design and Implementation (NSDI), 2022.}The advent of switches with programmable dataplanes has enabled the rapid development of new network functionality, as well as providing a platform for acceleration of a broad range of application-level functionality. However, existing switch hardware was not designed with application acceleration in mind, and thus applications requiring operations or datatypes not used in traditional network protocols must resort to expensive workarounds. Applications involving floating point data, including distributed training for machine learning and distributed query processing, are key examples.

In this paper, we propose \arch, a floating point representation designed to work efficiently in programmable switches. We first implement \arch on an Intel Tofino switch, but find that it has limitations that impact throughput and accuracy.
We then propose hardware changes to address these limitations based on the open-source Banzai switch architecture, and synthesize them in a 15-nm standard-cell library to demonstrate their feasibility. Finally, we use \arch to implement accelerators for training for machine learning and for query processing, and evaluate their performance on a switch implementing our changes using emulation. We find that \arch allows distributed training to use 25-75\% fewer CPU cores and provide up to 85.9\% better throughput in a CPU-constrained environment than SwitchML. For distributed query processing with floating point data, \arch enables up to 2.7$\times$ better throughput than Spark.

\end{abstract}

%% file: introduction.tex
\section{Introduction}
\label{sec:intro}

The rise of programmable network devices has 
transformed distributed systems design. Instead of simply moving data between servers using standard routing protocols, network devices can be programmed using domain-specific languages like P4~\cite{bosshart2014p4} and NPL~\cite{npl} to support new network functionality, 
such as congestion
control~\cite{211285}, load
balancing~\cite{10.1145/3098822.3098824,katta16:_hula}, and packet scheduling~\cite{10.1145/2934872.2934899}.
Commodity Ethernet switch ASICs with programmable data planes~\cite{barefoot,8526855,trident4} enable the execution of these programs at many terabits per second. 

While these capabilities were originally targeted at increasing network functionality, much recent work has explored their utility in accelerating application-level functionality as well.
Consensus
protocols~\cite{ports15:_desig_distr_system_using_approx,li16:_fast_replic_nopax,Dang.P4xos},
concurrency control~\cite{10.1145/3132747.3132751,jepsen18:_infin_resour_optim_concur_contr}, vector
addition~\cite{sapio17:_in_networ_comput_dumb_idea,sharp,sapio2019scaling}, query processing
operators~\cite{lerner2019case,10.1145/3230543.3230555}, and key-value
stores~\cite{10.1145/3132747.3132764,tokusashi18:_lake,li:_pegasus} have all been shown to benefit from this in-network computation~\cite{10.1145/3317550.3321439}.

However, an important class of applications has struggled to take advantage of in-network computation: those using floating point (FP) values.
These occur in two broadly-deployed datacenter applications: distributed training for machine learning, and distributed data processing systems. Since programmable switches were originally optimized for networking applications, their design includes basic support only for integer operations. Applications wanting to take advantage of in-network computation with floating point values have so far worked around this in one of three ways.

The first approach is to approximate floating point 
operations in software running on end-hosts. This is the approach taken by SwitchML~\cite{sapio2019scaling} as it sums gradient vectors as part of training deep neural networks. For each chunk of gradient vector elements, SwitchML executes a protocol that requires running code to convert between floating point 
and integer values on end hosts, as well as performing two rounds of communication. This protocol overhead is costly (see~\secref{sec:mlspeedup}).

The second approach is to build a switch ASIC that includes floating point hardware. This is the approach taken by the Mellanox Quantum switch~\cite{qm8700,sharpv2}. 
Dedicating chip resources for this purpose is expensive: we show
(\secref{sec:extensions}) that adding dedicated FPU hardware takes
more than $5\times$ the die area and power of integer ALUs. As a
result, this is not a general-purpose approach; it has only been taken
for InfiniBand switches, which have simpler routing designs and buffer
requirements than Ethernet switches, and hence have spare die area.
It also lacks flexibility: it is tied to specific operations on specific floating-point formats. New ML-specific numeric representations (\eg, FP16~\cite{micikevicius2017mixed,10.1145/3146347.3146358}, bfloat16~\cite{dean2012large,bfloat16,kalamkar2019study}, TF32~\cite{tf32}, and MSFP~\cite{msfp}) represent an area of ongoing innovation, and adding support for a new format requires developing and manufacturing a new ASIC -- an expensive and time-consuming endeavor. For example, it took four years for Mellanox to release its second version of switches with floating point support~\cite{10.5555/3018058.3018059,sharpv2}.

A related approach is to use FPGAs or other non-switch programmable devices to implement switch-like specialized accelerators~\cite{10.1145/3307650.3322259,7130,gebara-mlsys,flare}. While this yields a functional solution, the fine-grained programmability of a FPGA comes at the cost of power~\cite{10.1145/3302424.3303979} and area: for example, Xilinx's flagship FPGA supports $\sim$8~Tbps~\cite{versal} of Ethernet I/O, while the Intel Tofino 2, a regular programmable switch, supports 12.8~Tbps~\cite{barefoot2}. 

In this paper, we argue for a different approach. We propose \arch, which implements floating point computation as a P4 program running directly on a programmable switch. This is not straightforward: the multi-cycle nature of floating-point operations is at odds with the streaming-pipeline architecture common to P4-programmable switches today. To make it work, \arch breaks apart each floating point value into \textit{exponent} and \textit{signed mantissa} and stores them separately in different pipeline stages, decomposing the corresponding sub-operations appropriately to ensure correct execution. Rather than requiring specialized floating-point hardware, \arch repurposes network-oriented hardware elements in the switch pipeline to implement the sub-operations not supported by the switch's integer ALUs. 

\arch is a generic approach. We evaluate its feasibility on the Intel
Tofino~\cite{barefoot}, a commercially-available PISA switch. We
observe that constraints of the existing Tofino architecture present
obstacles to a full \arch implementation. We address this in two ways.
First, we introduce an approximate \arch design (\approxarch) that is
implementable on existing hardware, albeit with some precision and
throughput limitations. Second, we propose some simple and cheap hardware modifications, based on the open-source Banzai~\cite{10.1145/2934872.2934900} switch architecture, to enable high throughput and accuracy with \arch. We show that such enhancements are feasible in a 15-nm standard-cell library with minimal power, area, and timing cost relative to a baseline switch chip.

Finally, we implement accelerators for two use cases: distributed training for machine learning and distributed database queries, based on the recent SwitchML~\cite{sapio2019scaling} and Cheetah~\cite{10.1145/3318464.3389698} systems respectively. We use emulation to evaluate the performance of these use cases on a switch implementing our changes.
For distributed training, enhancing SwitchML with \arch (based on both regular FP32 and ML-specific FP16) allows it to use 25-75\% fewer CPU cores, giving up to an 85.9\% improvement in training throughput on CPU-limited configurations. 
For distributed query processing, a \arch-integrated version of Cheetah allows it to achieve comparable results for floating point queries as the original system does for integer queries -- a 2.3--2.7$\times$ speedup over Spark~\cite{spark}.

%% file: background.tex
\section{Background and Challenges}
\label{sec:back}

Conventional network switches are fixed-function, requiring redesign to add new features or support new protocols.
However, in today's era of software-defined networking~\cite{kreutz2014software}, rapidly evolving networking techniques and applications require new packet processing support.
Programmable switches, which allow the data plane behavior to be reconfigured, provide the necessary flexibility.
The RMT-based Protocol-Independent Switch Architecture (PISA)~\cite{10.1145/2486001.2486011} has emerged as the de facto standard for programmable switch architecture.

\begin{figure}[!t]
    \centering 
    \includegraphics[width=\linewidth]{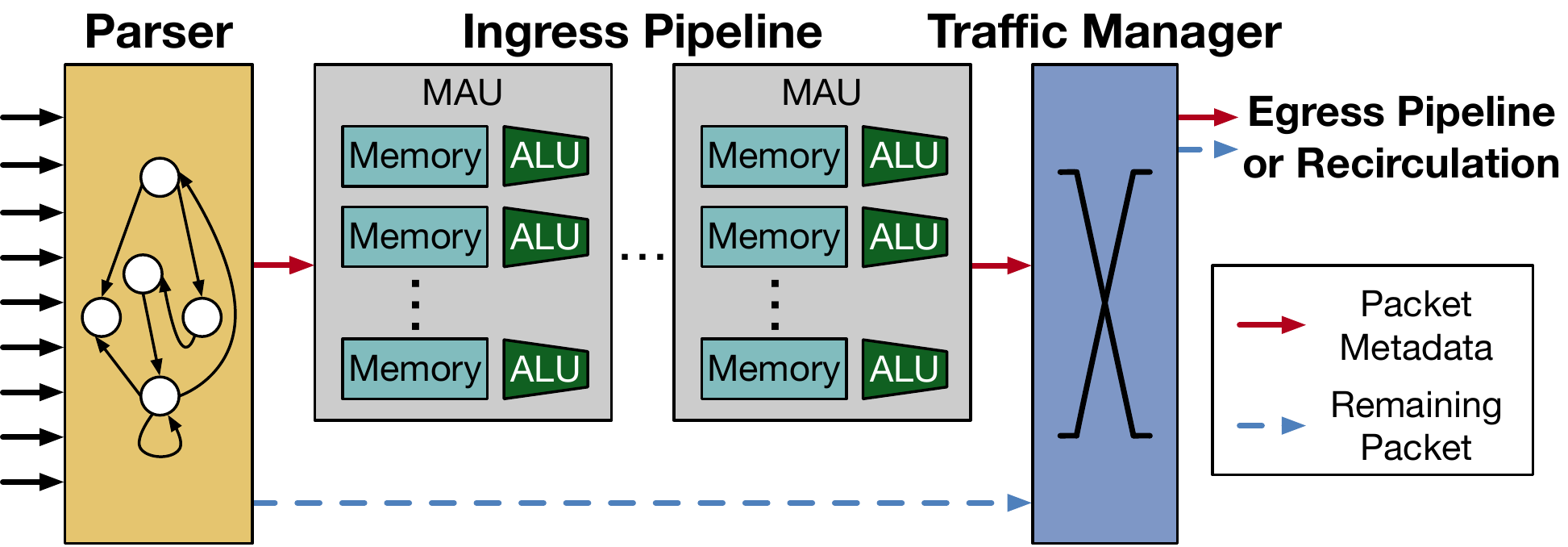}
    \caption{Basic PISA design.}
    \label{fig:pisa}
\end{figure}
\subsection{PISA}
\label{sec:pisa}

We 
depict the basic protocol-independent switch architecture design in \figref{fig:pisa}. 
The parser is a programmable state machine responsible for extracting user-specified fields of the inbound packet to per-packet metadata.\footnote{The remainder of the packet is passed through the pipeline, but cannot be matched or manipulated.}
The ingress pipeline consists of multiple cascaded match-action units (MAUs). 
Each MAU has some memory (SRAM and TCAM) and ALUs. 
It matches fields from the packet metadata against the memory to determine the corresponding action to be taken by the ALUs.
The ALUs support basic integer arithmetic and logic operations, and can be used to modify fields in the packet metadata.
They can also manipulate registers, which hold state that persists across different packets.
 
After going through the ingress pipeline, the packet
is routed to an egress port and queued by the traffic manager.
Before being output, it passes through an egress pipeline that has the same structure as the ingress pipeline, and the packet header and body are reassembled by the deparser.

Programmable switches following this architecture have become commercially available on commodity switches, thanks to reconfigurable switch silicon like the Intel (Barefoot) Tofino~\cite{barefoot} and Marvell XPliant~\cite{xpliant}.
A long line of research has showed how to use PISA switches to implement new networking protocols, offload network functions, and accelerate application-level logic~\cite{10.1145/3317550.3321439,hauser2021survey}.

\begin{figure*}[!t]
    \centering 
    \includegraphics[width=\linewidth]{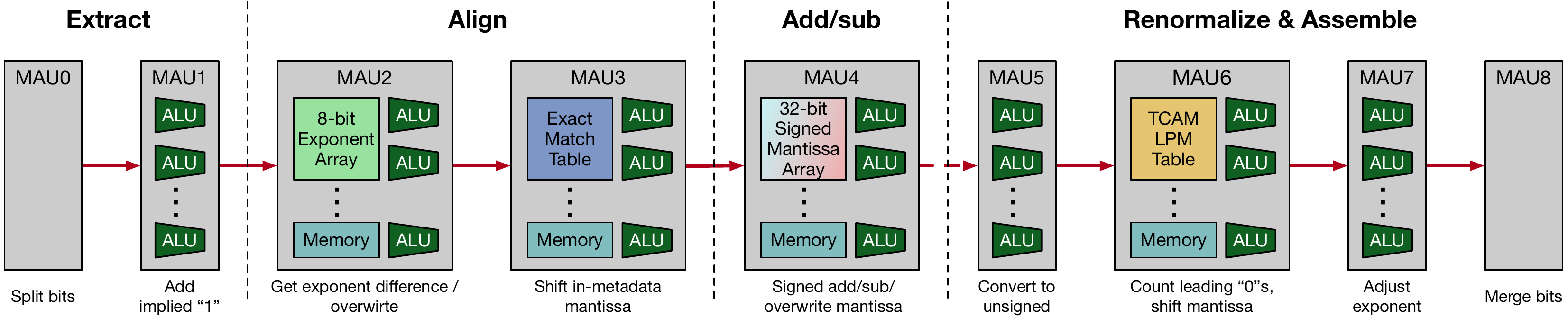}
    \caption{\arch dataflow. Only hardware components relevant to \arch are shown.}
    \label{fig:flow}
\end{figure*}
\subsection{Floating Point Overview}
\label{sec:add}
We describe the flow of the most common floating point operation in applications discussed in this paper -- addition -- here.
Note that subtraction is performed using the same process, and comparisons are typically implemented using subtraction. 
Regardless of specific widths, floating point values are represented with three parts: 1-bit sign, $n$-bit exponent, and $m$-bit mantissa.  
Typically, a floating point number is represented in normalized form: the mantissa value is in the range of $[1,2)$, \ie, it begins with a leading ``1'' bit (which can be omitted, \ie, ``implied 1'').
A floating point addition $C=A+B$ is performed using a five-step process: 
(We assume here that $\text{abs(}A\text{)} \le \text{abs(}B\text{)}$.)

\niparagraph{Extract.} The three parts of $A$ and $B$ are extracted from the packed data. 
The implied ``1'' in the packed mantissa is expressed explicitly. 

\niparagraph{Align.} The two mantissas are aligned to represent values at the same magnitude. 
Specifically, $\textit{mantissa}_A$ (the smaller one) is right-shifted by $\textit{exponent}_A-\textit{exponent}_B$ bits.

\niparagraph{Add/subtract.} Now that the two mantissas are aligned, they are added or subtracted, depending on sign: $\textit{mantissa}_C = \textit{mantissa}_B \pm \textit{mantissa}_A$.

\niparagraph{Renormalize.} The result is scaled so that the mantissa is in the range of $[1,2)$.
This is achieved by counting the leading ``0'' bits and left or right shifting $\textit{mantissa}_C$ accordingly, then adjusting $\textit{exponent}_C$ by the corresponding value.

\niparagraph{Round and Assemble.} Finally, the three parts of $C$ are packed into a single value.
The implied leading ``1'' of $\textit{mantissa}_C$ is stripped. If more mantissa bits are available than can be represented in the packed format, the mantissa is rounded.

\subsection{Challenges}
\label{sec:challenges}

Current PISA architectures do not natively support any floating point operations.
This is no surprise, considering that they were designed for packet processing, and floating point support is expensive.
FPUs have much larger power and area costs than integer ALUs~\cite{4039591,1374989,10.1145/1669112.1669172}, and the complex floating point addition procedure (\secref{sec:add}) takes multiple cycles and thus introduces timing constraints.

This paper asks if we can build floating point addition operations on a commodity PISA architecture.
Intuitively, it should be possible to decompose the canonical addition procedure and span it across multiple pipeline stages.
However, we observe that this leads to two challenges.

First, registers are associated with specific pipeline stages, and can only be accessed from that stage. 
That is, each register can only be accessed once per packet, and data dependencies cannot ``go backwards'' to an earlier stage.\footnote{Recirculating an entire packet is an exception. However, it is costly and bandwidth constrained.}
This poses a problem for applications, like in-network aggregation, that wish to maintain and update floating point state: it is not possible, for example, to perform the add-mantissa and renormalize steps in different pipeline stages.

Second, the available ALU operations may not be sufficient to implement all the operations necessary to implement floating point addition. For instance, on a CPU, the renormalization step might use a count-leading-zeros instruction (\eg, \texttt{lzcnt} on x86), but we know of no PISA switch with such an instruction.

Hence, we must develop a PISA-friendly, decentralized (multi-stage) approach for floating point addition.

%% file: design.tex

\section{\arch Design}
\label{sec:design}

How can we implement floating point operations on PISA architectures, given the challenges described above?
We propose a design, \arch, based on a new floating point representation and a mapping of its operations to PISA pipelines, as shown in \figref{fig:flow}.
In this section, we describe the basic \arch approach in the context of an abstract PISA pipeline; \secref{sec:arch} discusses additional challenges that occur when implementing it on existing PISA architectures.

\arch has three key ideas:

\niparagraph{Decoupled exponent and mantissa operations.}
\arch processes operations on the exponent and (signed) mantissa components of FP values separately, and internally stores them in separate registers.
This decoupling allows them to be processed by different pipeline stages.

\niparagraph{Delayed renormalization.}
Second, \arch does not require intermediate values to be renormalized on every operation.
That is, in a SwitchML-like~\cite{sapio2019scaling} aggregation workflow, values from each client are added to an accumulator whose value is not renormalized until the final result is output. This is based on two observations about FP renormalization.
First, renormalization does not affect the correctness of floating point operations.
Scaling the mantissa to place the leading ``1'' in its correct location is needed to produce an output value in canonical format, but a denormalized form can equally represent the same arithmetic value.
Second, renormalization introduces data dependencies between the mantissa and exponent components, which makes it challenging to fit into a PISA pipeline.
In particular, renormalization requires the exponent to be adjusted based on the computed mantissa, whose computation itself depends on the exponent -- a circular data dependency that cannot be represented in a single pipeline traversal.
To avoid this, when we read from the accumulator, we read the denormalized value, and normalize it just before sending out the final result. We do not store the normalized value back into the accumulator.

\niparagraph{Extra bits in mantissa register.}
PISA architectures commonly have registers with limited bit widths: 8-, 16-, or 32-bit registers are common; on the other hand, FP values commonly have mantissas with smaller bitwdith.
We take advantage of this difference in two ways. First, we can use bits to the right of the mantissa as guard bits to aid in rounding, as is common in standard FP ALUs. Second, we can use bits to the left of the mantissa to avoid overflow when summing multiple values with similar exponents. When we add two values with mantissas that are all ones, the addition simply carries into the bits to the left of the mantissa.

In this section, we use IEEE~754 FP32 -- which has a 1-bit sign, 23-bit mantissa, and 8-bit exponent --  as an example to demonstrate \arch design.
Other FP formats with different widths can also be supported.
\figref{fig:flow} shows \arch's dataflow.

\begin{figure}[!t]
    \centering 
    \includegraphics[width=\linewidth]{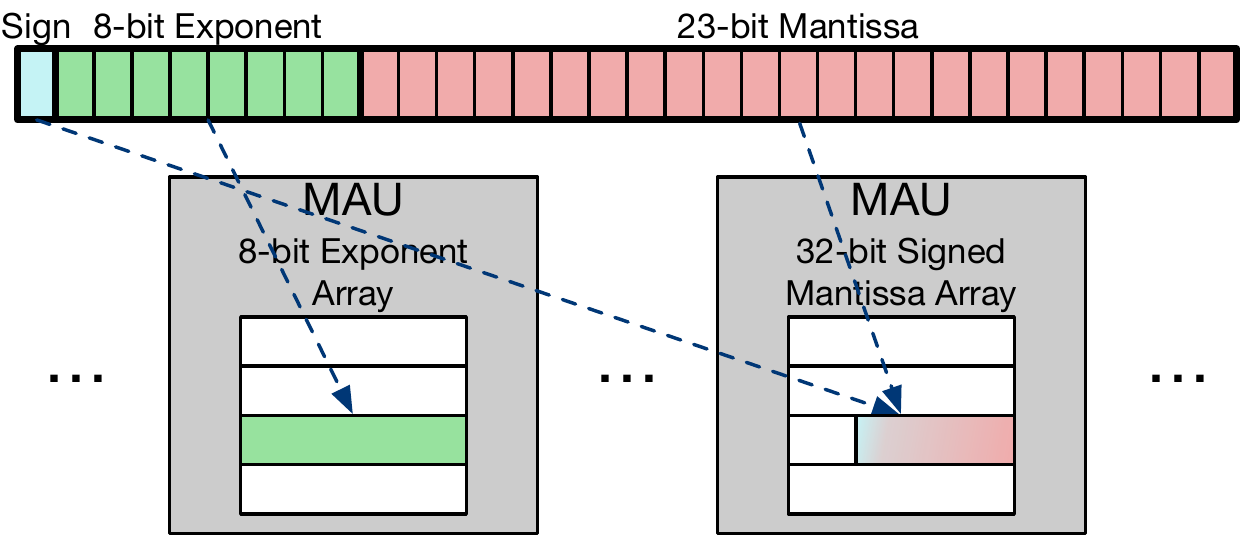}
    \caption{\arch's representation of FP32 in the switch.}
    \label{fig:format}
\end{figure}
\subsection{Representing FP in PISA}

To meet the constraints of PISA, \arch splits the storage of FP values using the representation shown in \figref{fig:format}.
The exponent field is stored in an 8-bit-wide register array. 
The 23-bit mantissa is stored, right-aligned, in a 32-bit register.  
To unify signs and addition/subtraction operations, we store the mantissa in two's-complement signed representation.

\arch needs more memory space to store a floating point number (\eg, 8+32=40 bits for a FP32 number). However, we argue that this will not significantly reduce the efficiency of \arch since exponent and mantissa have to be stored in different MAUs anyway. Hence, the per-MAU parallelism of floating point operations will not be affected.

\subsection{Performing FP operations in PISA}

\begin{figure}[t]
  \centering
  \includegraphics[width=\linewidth]{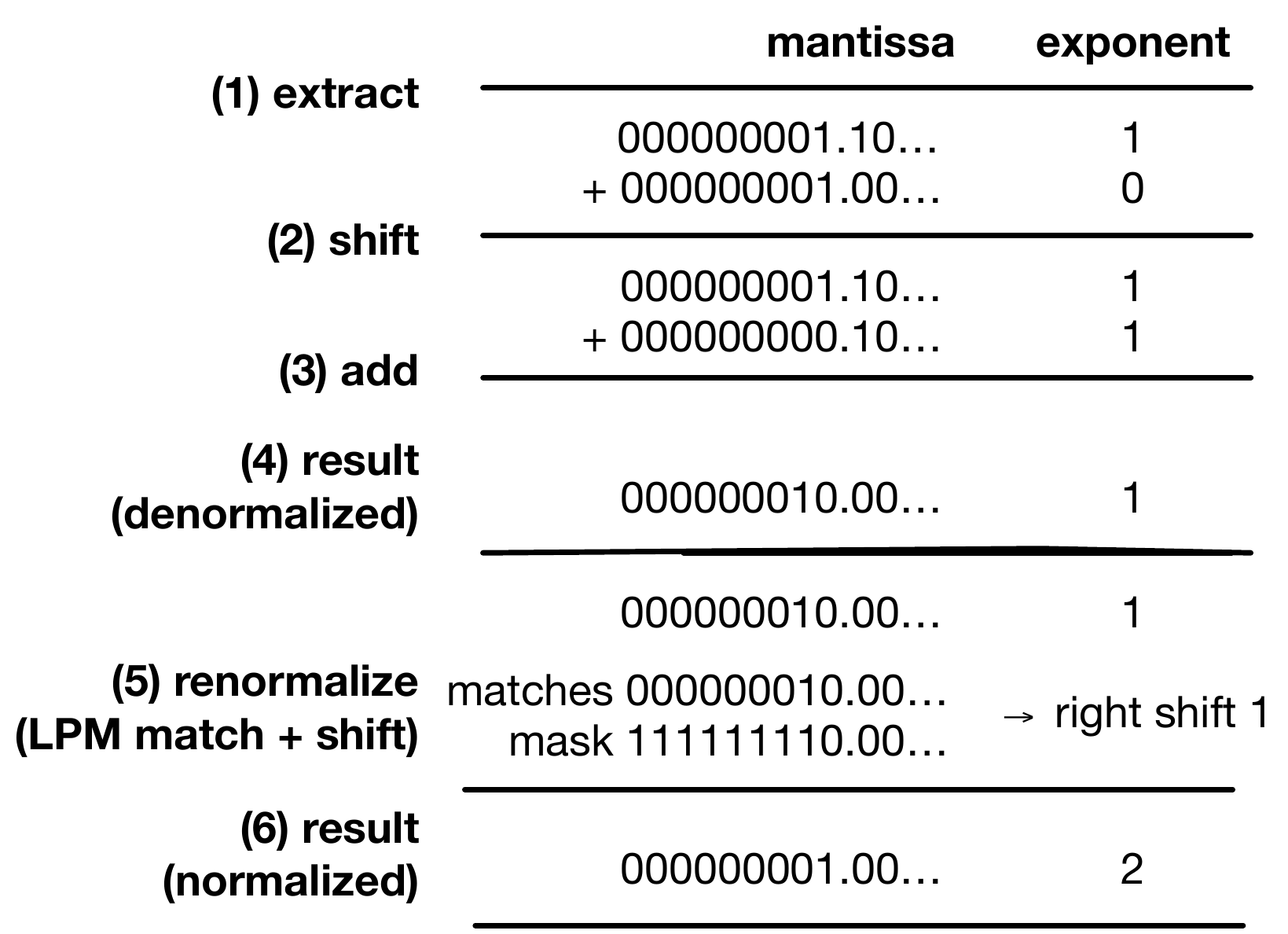}
  \caption{Example of \arch addition: computing the sum of 3.0
    (0b$1.1 \times 2^1$) and 1.0 (0b$1 \times 2^0$). Computation is
    done using a 32-bit mantissa; 21 trailing zero bits are elided.}
  \label{fig:example}
\end{figure}
By delaying renormalization until the output phase and storing exponents and mantissas separately, \arch makes it possible to adapt the standard extract-align-add-renormalize-assemble FP addition flow to a PISA pipeline.
\figref{fig:flow} shows the mapping of functionality to MAUs.
We use a running example (\figref{fig:example}) where an input of
$1.0$ is added to a register containing the value $3.0$.

\niparagraph{Extract.}
The first stages extract the exponent and mantissa from a FP32 value in the input packet into separate metadata registers (MAU0), then add the implied ``1'' to the extracted mantissa field (MAU1). 
The decoded values are shown in \figref{fig:example} step (1). 

\niparagraph{Align.}
\arch then compares the provided exponent value with the one stored in
memory in MAU2.
This updates the exponent and determines which of the two operands's mantissa must be right-shifted and by how much.
The right shift itself is performed for the metadata value by MAU3, and for the memory value by MAU4 (where the mantissa register is located).
In \figref{fig:example} step (2), 1.0 is shifted right to be
expressed as $0.1 \times 2^1$

\niparagraph{Add.}
In addition to shifting the mantissa of the in-memory value, MAU4 performs the mantissa addition itself.
Depending on the sign bit, it either adds or subtracts the shifted
mantissa value generated in the previous stage from the stored
mantissa value (step (3) in \figref{fig:example}).
The resulting mantissa value replaces the previous stored mantissa.

Note that MAU4 is used both to perform the right shift of the stored mantissa and its addition.
This is a necessity because the PISA architecture can only update a given register from one stage.
Existing implementations may not be able to perform both operations with a single stateful ALU; we discuss how to extend them or how to work around this limitation in \secref{sec:arch}.

\smallskip

At the end of this process, the exponent and mantissa registers contain the result of the addition, but may not be in normalized form.
For example, in step (4) of \autoref{fig:example}, the registers
store the value 0b$10.0 \times 2^1$. This is indeed a valid
representation of the result $4.0$, but is not in normalized form
because the mantissa has more than one digit to the left of the binary point.

\begin{figure}[t]
      \centering
      \includegraphics[width=\linewidth]{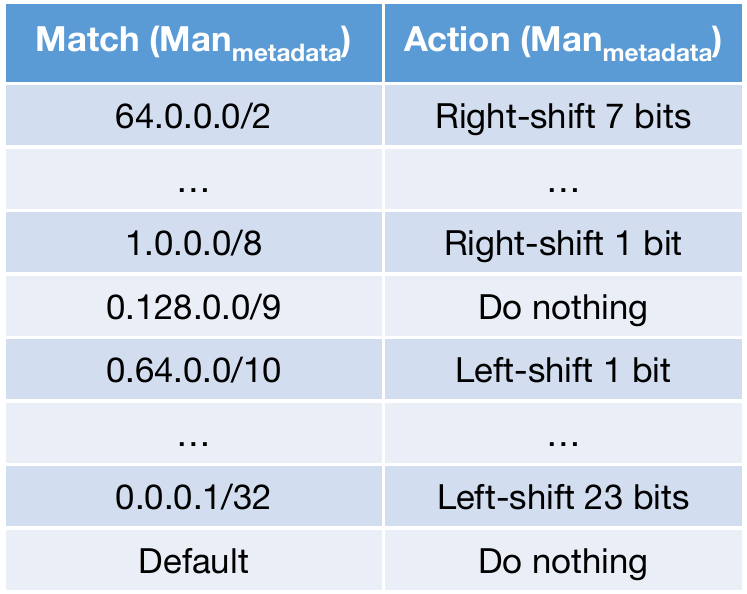}
     \caption{LPM match-action table (MAU6) in \arch design.}
     \label{fig:table2}
\end{figure}

\niparagraph{Renormalize and Assemble.}
\arch delays renormalization: it does not renormalize the intermediate value stored in registers, but only when the result is to be output.
Thus, multiple additions can be performed before renormalization.
This offers two benefits.
As mentioned before, it eliminates the need to adjust the exponent stored in memory after calculating the mantissa, avoiding a data dependency.
Second, since the renormalization and assembly steps are stateless, we can place them in the (normally underutilized) egress pipeline, making more efficient use of resources.

The renormalization process itself is performed in four steps.
The aggregated mantissa is first converted from its two's complement signed representation to unsigned value and sign (MAU5).
\arch then counts the number of leading zeros and shifts the mantissa value accordingly, in order to place the leading ``1'' bit in the right location (MAU6).

Because no PISA switches support a count-leading-zeros operation, \arch{} exploits a TCAM-based longest prefix match (LPM) table -- commonly used in IP routing -- to implement this function. Specifically, we construct a LPM table where each entry has an IP address with only the $i$th bit set, and a netmask that matches the first $i$ bits. A match indicates that the mantissa has $i-1$ leading zeros. This is used to select the right shift action that places the leading 1 in its canonical location (bit 24 for FP32).
In the example, the leading ``1'' is located using a match,
whose bitwise representation is shown in step (5), which corresponds
to the CIDR address 0.128.0.0/9; the lookup table
(\figref{fig:table2}) indicates that the mantissa should be shifted
right by 1.
The exponent is adjusted also according to the leading zeros' count
(in MAU7) -- here, incremented by 1.  
This gives a normalized result; all that remains is to merge the sign,
exponent, and lower 23 bit of the 32-bit mantissa fields (in MAU8) to
put it in FP32 format.

\subsection{Discussion}

\niparagraph{Overflow.} 
The denormalized representation has the potential to overflow if similar values are added many times. With a signed register size of 32 bits and a mantissa size of 24 bits, there are 7 bits to the left of the mantissa available for holding overflows. This is sufficient to represent 128 additions of values with the maximum mantissa with the same exponent -- an extreme case -- into a single register without overflow. However, for the use cases described later in the paper, the number of operations per register is equivalent to the number of nodes in the distributed system. If overflow occurs, it can be detected and signaled to the user, who can handle it in an application-specific way.

\niparagraph{Other FP formats.} \arch can be trivially modified to
support floating point formats with different exponent and mantissa
width (\eg FP16, which we evaluate in \secref{sec:ml}). Likewise,
block floating point formats, where multiple values share one exponent~\cite{msfp}, can be supported by replicating the exponent register.

\niparagraph{Rounding and reproducibility.} As described, \arch
provides round-to-negative-infinity semantics, and reproducible
results (though not necessarily consistent with IEEE 754). These
semantics are acceptable for the applications we study; a further
discussion is in Appendix~\ref{sec:advancedops}.

%% file: arch.tex
\section{Realizing \arch on PISA Architectures}
\label{sec:arch}

The previous section shows how \arch can map floating point operations to an abstract PISA architecture.
Actual PISA implementations may have restrictions on MAU operations.
We have implemented \arch in P4 for the Barefoot Tofino switch.
In doing so, we encountered several architectural limitations (\secref{sec:impl_c}).
We show that simple architectural extensions, which can be implemented with minimal power and chip area cost, can resolve these limitations and enable a full \arch implementation (\secref{sec:extensions}).
Alternatively, we describe an approximate approach, \approxarch, which
works around these limitations to implement a variant of \arch for the existing Tofino architecture, albeit with tradeoffs in accuracy and resource utilization (\secref{sec:approx}).

\subsection{Challenges}
\label{sec:impl_c}

We implement \arch addition in the P4 language~\cite{bosshart2014p4} ($\sim 580$ LoC) in a modularized manner (\ie, one FP addition per module) and compile it to Intel (Barefoot) Tofino ASIC~\cite{barefoot}. 

Using this implementation, we identify three limitations of the the current Tofino hardware that impact the functionality and efficiency of our FP operations.

\niparagraph{Resource utilization of shift operations.}
In general, multiple \arch modules can be deployed in parallel, sharing the same pipeline stages and overlapping with each other.
For many applications, performing as many operations per packet as possible is essential to achieving high performance~\cite{sapio2019scaling}.
Unfortunately, the current Tofino architecture can only accommodate one \arch module in its ingress pipeline, \ie, only one FP addition can be performed per packet.

After analyzing the resource utilization (Appendix~\ref{sec:resource}), we observe that the main source of overhead is performing shift operations.
Specifically, \arch needs to shift fields by a variable number of bits, in order to implement the alignment and renormalization stages.
However, the Tofino ALUs can only perform shift operations with a fixed shift distance, specified as an immediate.
While it is possible to emulate a variable-length shift operation with
the current functionality, doing so is resource intensive. In
particular, per-stage VLIW instruction utilization prevents multiple \arch instances from sharing pipeline stages.

\niparagraph{Lack of atomic shift-and-add.}
One of the pipeline stages in the abstract design (MAU4 in \figref{fig:flow}) must perform two operations: right-shifting the stored mantissa to align it with the value being added, and performing the mantissa addition.
Both are stateful operations on the mantissa register, so they must be performed by the same stage's ALU.
However, the Tofino's ALUs cannot perform \emph{both} a shift and an add operation.
In \secref{sec:approx}, we show how to work around this limitation by left-shifting the other mantissa value (from the packet metadata) instead; this allows the \arch design to be implemented on the existing Tofino architecture, but can lead to numerical error for some workloads.

\niparagraph{Endianness conversion.}
While hardly unique to \arch, endianness conversion is a non-trivial source of overhead for \arch applications.
Network devices interpret values in network byte order (big-endian), whereas most general-purpose CPUs are little-endian.
To identify and process the data correctly in the switch, endianness conversion is necessary. 
Traditional networking applications only need to convert byte order for headers, which are relatively small.
For data-intensive in-switch applications, byte order conversion for the full payload can have high overhead. 
While the Tofino has functional units that can do this conversion, they are not plentiful enough to convert full payloads, and thus the conversion must be done on end hosts.

\begin{figure}[!t]
    \centering 
    \includegraphics[width=\linewidth]{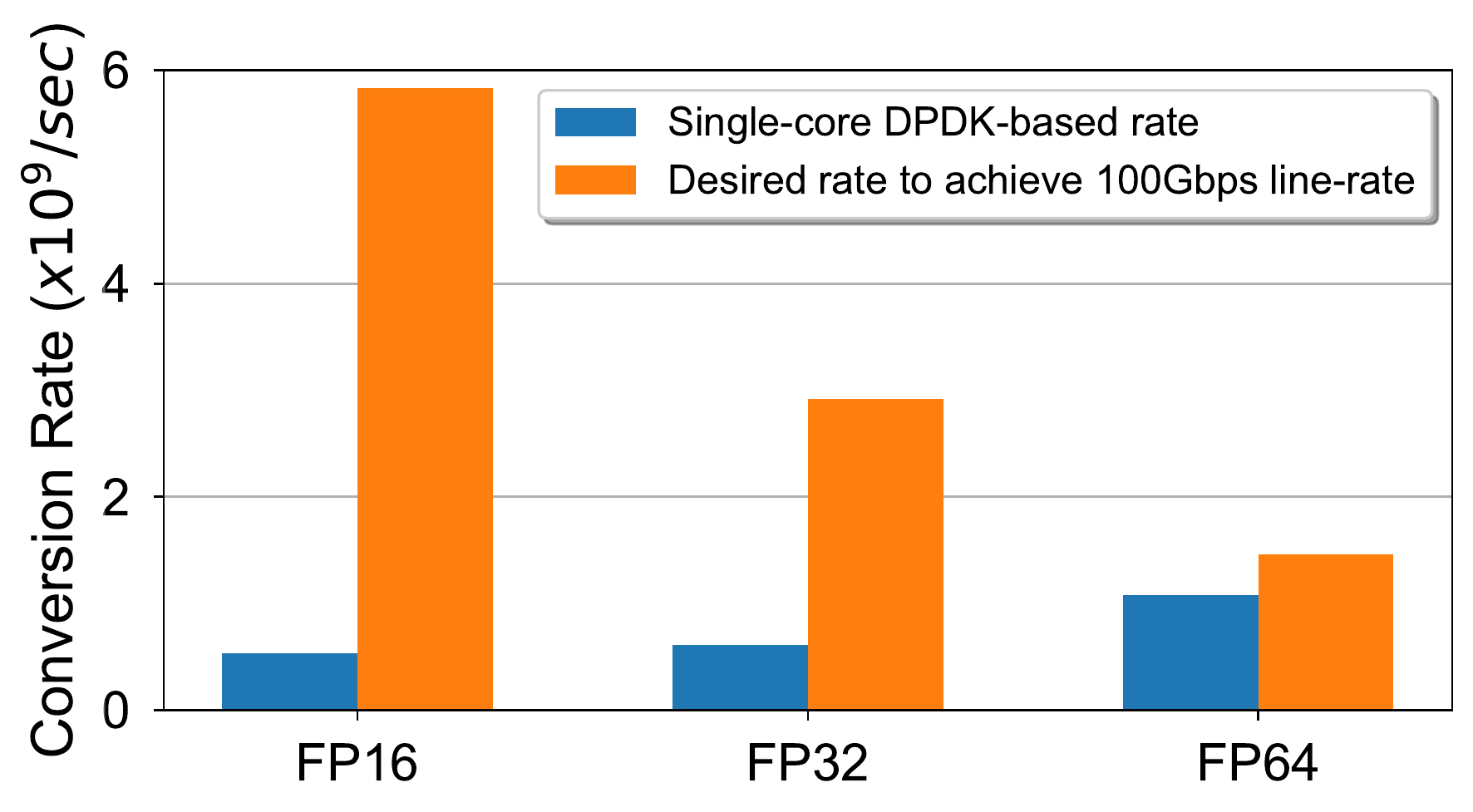}
    \caption{Endianness conversion rate that a core can achieve and that is desired to achieve 100Gbps line-rate.}
    \label{fig:endian}
\end{figure}

To quantify the overhead, we test how rapidly a single x86 core (running at 2.3GHz) can perform endianness conversion for different FP formats, using DPDK's highly-optimized APIs with ``O3'' optimization.
\figref{fig:endian} compares the measured results with the rate needed to achieve line-rate conversion at 100~Gbps.
The gap is large, particularly for lower-precision values. 
In particular, to reach 100 Gbps for FP16, one will need at least 11 (\ie, $\lceil \text{desired rate} /  \text{single-core rate}\rceil $) cores.
Hence, the high overhead of endianness conversion will lead to either low network throughput or extra CPU or GPU utilization. 
In many applications, these resources are not free; for instance, in DNN training, CPUs are often busy with data preprocessing.

\subsection{PISA Architectural Extensions}
\label{sec:extensions}

To avoid these problems, we propose to extend the PISA architecture with some additional support. We show that the cost of these additions is low by extending the Banzai switch architecture model~\cite{10.1145/2934872.2934900} and demonstrating that the increase in chip area, power, and timing budget is not significant.

\niparagraph{2-operand shift instruction.} We propose to enhance the existing shifter by allowing the shift distance operand to come from metadata instead of an immediate.
The proposed instruction format is \texttt{shl/shr reg.distance, reg.value}.
This little-effort enhancement will significantly improve the resource efficiency of \arch, since the shifter can directly take the table match result as operand, and two instructions (left- and right-shift) can handle all the cases. 
\begin{table}[!tb]
  \centering
  \caption{Stateless ALU and stateful RAW/RSAW unit  areas and minimum critical-path delays in FreePDK15 library. Each of the compiler targets contains 300 instances of one of the ALUs. Power and area are evaluated at 1~GHz frequency target.}
  \vspace{-1ex}
\scriptsize
  \label{tab:shift}
  \begin{tabular}{m{2.3cm}m{0.6cm}m{0.6cm}m{0.6cm}m{0.6cm}m{0.6cm}}
    \toprule
    &  \bf Default ALU & \bf \arch ALU &  \bf Default RAW & \bf \arch RSAW & \bf ALU+ FPU\\
    \midrule
  Dynamic power ($\mu W$) & 594.2 & 669.4& 637.6 & 721.1 & 3590.6\\ 
  Leakage power ($\mu W$) & 18.6 & 22.8& 16.8 & 22.1 & 109.8\\
  Area ($\mu m^2$)        & 505.4 & 618.6 & 468.8 & 633.0 &3837.7\\
  Min Delay (ps) & 133 & 135 & 133 & 151 & 136\\
  \bottomrule
  \end{tabular}
\end{table}

\niparagraph{Combined shift+add operation in one stage.} 
If the switch can support an atomic ``shift+add'' operation on a register in a single stage, we will be able to swap the mantissa, with no compromise of potential error.

\niparagraph{In-parser hardware-based endianness conversion.}
Endianness conversion in the hardware is straightforward and cheap -- pure combinational logic shuffling the wires. 
We propose a simple enhancement to the switch's parser and deparser to implement this.
Specifically, we propose a P4 type annotation \texttt{@convert\_endianness}, applied to entire headers, that indicates to the compiler that the parser and deparser should convert the header fields' endianness as they enter and leave the pipeline.
The parser will store the corresponding result to the metadata along with a implicit tag bit adjacent to the header's valid bit.
When the packet is re-assembled, the deparser will check this tag bit to determine the byte order to be emitted.

To evaluate the cost of the first two changes (the last change has
near-zero cost), we modify the open-source Banzai~\cite{10.1145/2934872.2934900} switch architecture, a PISA-like design.
We modify the Verilog code for Banzai's ALU to support our proposed shift instruction and synthesize it using Synopsys Design Compiler~\cite{dc} with the FreePDK 15nm FinFET standard-cell library~\cite{10.1145/2717764.2717783}, a technology node similar to that used by the Tofino. 
We first check whether the design can operate at 1~GHz, evaluate its power and area, and then search the minimum critical-path delay of each design to find the impact of our modifcation on timing. 
As the results in \tabref{tab:shift} show, an enhanced ALU may use 13.0\% more power and 22.4\% more area than the original ALU, while slightly increasing the minimum delay. 
The overhead mainly comes from connecting and storing the second operand in the shifter. 
We implement a stateful read-shift-add-write (RSAW) unit based on Banzai's atomic predicated read-add-write (RAW) unit. 
The synthesis results in \tabref{tab:shift} demonstrate that the RSAW unit uses 13.6\% more power and 35.0\% more area than the regular RAW unit.
In terms of minimum delay, RASW is 13.5\% longer than RAW, but still far from the 1ns bound at 1~GHz.
Banzai provides implementations only for the functional units, not for the entire switch chip, so we are unable to directly evaluate the impact of our modifications on the full chip design. However, prior work suggests that ALUs take up only a small portion (\ie, $\sim10\%$) of the power/area budget for the entire chip~\cite{10.1145/2486001.2486011}; from this we infer that our modifications would have negligible impact.
In other words, this hardware enhancement is feasible today, and is unlikely to become a bottleneck in future hardware generations.

Finally, to compare our approach with one that includes specialized floating-point units (like the Mellanox Quantum switch~\cite{qm8700,sharpv2}), we synthesize an ALU that includes a hard floating point unit. The ALU+FPU column in \tabref{tab:shift} shows the result: the hard FPU is more than five times larger and more power hungry than either the default ALU or the \arch ALU. Its high area and leakage power are costs that must be paid even when the FPU is not in use, making it challenging for a switch chip including these features to be competitive with ordinary switches in terms of efficiency, and forcing vendors to maintain separate specialized switch designs for different applications. Conversely, the \arch approach allows the same ALUs to support both floating-point and non-floating-point computations, enabling a single switch chip design to support both floating-point and non-floating-point workloads efficiently.

\subsection{\approxarch: \arch on Existing Architectures}
\label{sec:approx}
The architectural changes described above allow us to implement the full \arch approach.
We additionally want a solution that allows \arch to run on existing Tofino switches.
Achieving this requires addressing the shift-and-add limitation.
(The other two, while important, impact only resource utilization.)
We provide a way to approximate \arch on existing switches by avoiding the problematic shift.
This approximation, which we call \approxarch, can lead to inaccuracies for certain patterns of inputs, though we show later that it is not a problem for some applications, including in-network aggregation for ML training workloads (\secref{sec:ml}).

Recall that the problem arises because the alignment phase may require shifting the in-memory mantissa value to align it with the value to be added, which conflicts with the need to perform addition on the same value.
Note that this is not a problem when the in-memory value has a larger exponent than the in-metadata value, as only the smaller of the two is right shifted.
Taking advantage of \arch's tolerance for denormalized representations, \approxarch \emph{always shifts the in-metadata mantissa} rather than the in-memory value.
That is, if the in-metadata value is larger than the in-memory value, we keep the exponent unchanged and left-shift the in-metadata mantissa. 

This approach works, within a certain range, because \arch internally uses wider registers for the mantissa than the basic floating point representation.
For FP32, IEEE~754 uses a 23-bit mantissa, while \arch stores it in a 32-bit register.
This gives 7 bits of \emph{headroom}, after accounting for the implicit 1-bit and the sign bit.
If the value being added is \emph{much} larger than the in-memory value, i.e., its magnitude is greater by a ratio of more than $2^7 = 128$, the headroom would be exceeded.
Instead, we detect this case during the exponent comparison (MAU2 in \figref{fig:flow}) and replace the in-memory value entirely with the in-metadata one.
Doing so introduces numeric error in the low-order bits.

The \approxarch variant can be deployed to a commodity Tofino switch today (as we do in our evaluation).
As described above, it can introduce numeric error (which we call ``overwrite'' error).
However, the error only occurs when input values vary widely in magnitude, and is bounded by the difference between headroom and mantissa width.
For some applications, this approximation poses little difficulty: as we demonstrate in \secref{sec:ml}, ML model training gradients generally have a relatively narrow exponent range, and the workload is in any event resilient to small inaccuracies.
For others, it may be more problematic.
In these cases, the architectural modifications of \secref{sec:extensions} will be needed.

%% file: ml.tex
\section{Case Study: Distributed ML Training}
\label{sec:ml}

As the model and dataset sizes have increased for ML training jobs, large-scale distributed training has become increasingly important~\cite{199317,dean2012large,ho2013more,186212,10.5555/2685048.2685095,moritz2015sparknet,iandola2016firecaffe,10.1145/3419111.3421307,10.1145/3341301.3359642,227623,258953,265013,10.1145/2465351.2465371}. 
In this paper, we focus specifically on data-parallel training, a common approach to distributed training.\footnote{Other parallel modes, like model-parallel, may also benefit from what is discussed in this work, but we do not explore them here.}
In data-parallel training, the  dataset is partitioned to multiple worker machines, each with a replica of the  model. 
In a training iteration, each machine performs learning on its local dataset and model, generating gradient vectors. 
These gradient vectors are then used to update the weights that make up the model.
Modern supervised ML typically employs stochastic gradient descent (SGD)~\cite{nemirovsky1983problem,nemirovski2009robust,robbins1951stochastic} or its variants as the optimizer for iterative training. 
In general, the core operation of SGD is as follows: 
\begin{subequations}
	\label{eq:sgd}
\begin{align*}
	\displaystyle \textit{weight}_{(next)} &=  \textit{weight}_{(current)} - \textit{learning\_rate} \cdot \textit{gradient}_{(current)},
\end{align*}
\end{subequations}
where $gradient_{current}$ is the element-wise mean of all the local gradient vectors produced by each worker. Computing this mean requires summing (or aggregating) gradient vectors from all workers.

Prior work has observed that, as the number of workers and the size of the model grows, communication costs -- specifically, the gradient aggregation procedure -- increasingly become a bottleneck in distributed training~\cite{10.1145/3307650.3322259,10.1109/MICRO.2018.00023,10.1145/3405671.3405810}.
Gradient aggregation can be viewed as an ``all-reduce'' collective operation, a familiar concept from the HPC world -- the gradient vectors are \textbf{gathered} from all worker machines,  \textbf{reduced} to one vector, and \textbf{sent back} to all worker machines. 
It is traditionally implemented either using a parameter server~\cite{10.5555/2685048.2685095} or a distributed reduction protocol like ring all-reduce~\cite{intercom,Patarasuk2009117}.

In-network aggregation has been proposed as a promising way to accelerate this collective operation, and thus distributed training~\cite{10.1145/3307650.3322259,9138924,10.5555/3018058.3018059,sharpv2,adiga2002overview,sapio2019scaling,265053,gebara-mlsys}. 
In-network aggregation performs the ``reduce'' (\ie, sum) step of all-reduce in a network switch on the fly.
This offers higher throughput and lower latency than a parameter server approach, where both the network link and host-side network stack can become bottlenecks. 
Compared to ring-based and other distributed all-reduce algorithms, in-network aggregation requires exchanging fewer messages, again reducing latency and network usage.

PISA switches are well suited for, and have been used for, implementing in-network aggregation without specialized hardware. 
A major challenge, however, is the lack of floating point support.
The recent state-of-the-art in-network aggregation work, SwitchML~\cite{sapio2019scaling}, works around this by quantizing floating point values at end hosts so that the PISA switch only operates on fixed-point values.
While this quantization approach has been shown not to impact accuracy~\cite{sapio2019scaling}, we show that it harms performance.
In particular, quantization and format conversion requires significant CPU overhead on the worker hosts.
Computing the scaling factor to use for each block also requires an additional network round trip.
Both costs could be avoided if the switch could operate on floating point values directly.

\subsection{Characteristics of Training Gradients}
\label{sec:num}
The gradient aggregation workload has some common numerical characteristics that make it well suited for in-network aggregation with \arch. 
In particular, \arch can be used with existing Tofino switches using
the \approxarch approximation (\secref{sec:approx}); the resulting numerical error is rare and (as we demonstrate) has no impact on training accuracy.

\niparagraph{High aggregation parallelism.} In general, for each training iteration, the entire gradient vector corresponding to the training model needs to be aggregated from all worker machines.
These vectors can range from several MBs to GBs.
Aggregation is just vector addition; this element-wise summation provides ample parallelism.

\niparagraph{Vector-wise distribution.} As studied in INCEPTIONN~\cite{10.1109/MICRO.2018.00023}, gradient values in each vector largely fall in the narrow range of $[-1,1]$, and most are close to ``0''.

\begin{figure}[t]
    \begin{subfigure}[b]{0.32\linewidth}
      \includegraphics[width=\textwidth]{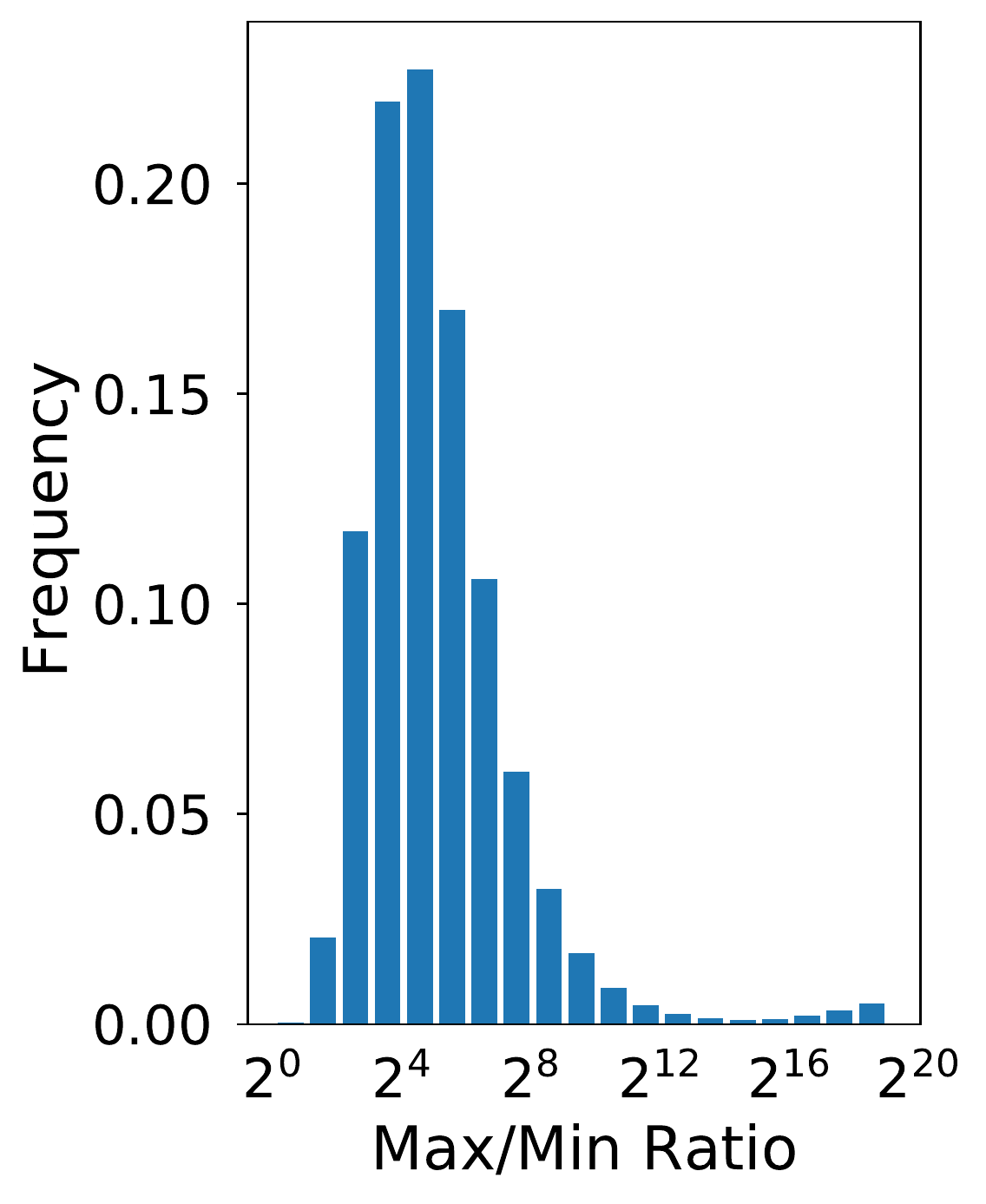}
      \caption{VGG\\(CIFAR-10).} 
      \label{fig:data1-1}
    \end{subfigure}
     \hfill
    \begin{subfigure}[b]{0.32\linewidth}
      \includegraphics[width=\textwidth]{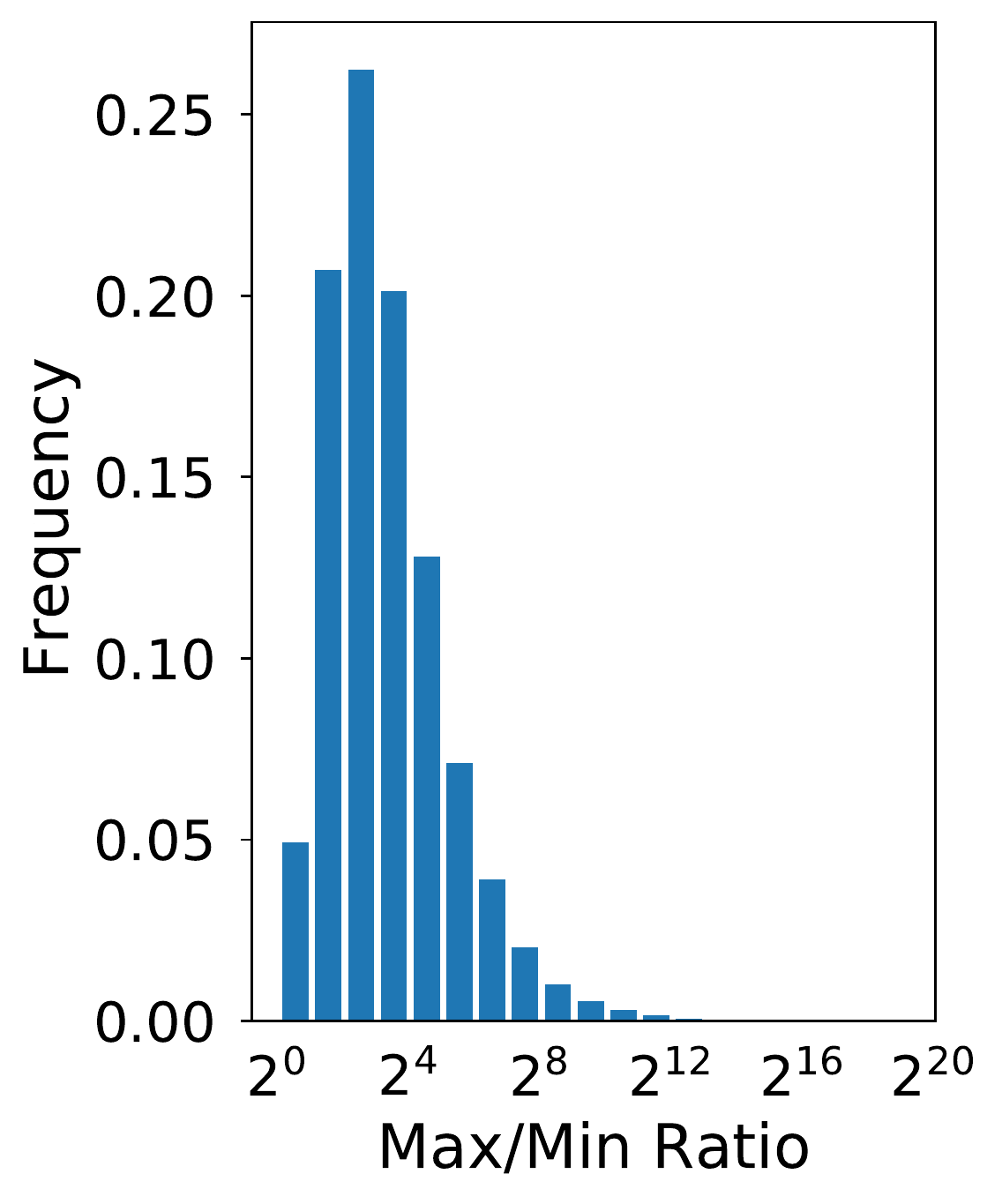}
        \caption{DeepLight\\(Criteo 1TB).}
      \label{fig:data1-2}
    \end{subfigure}
     \hfill
    \begin{subfigure}[b]{0.32\linewidth}
      \includegraphics[width=\textwidth]{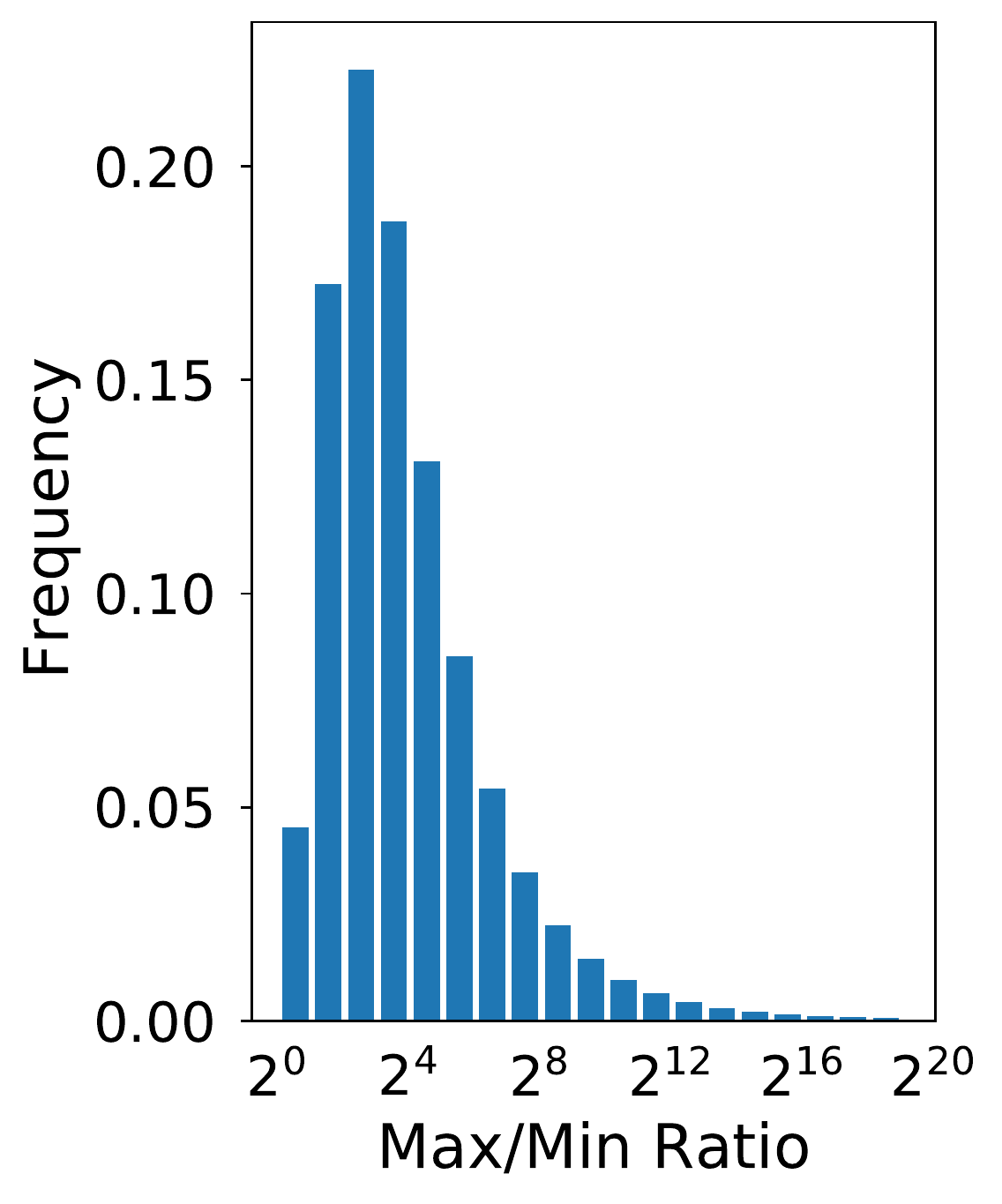}
      \caption{LSTM\\(GBW).}
      \label{fig:data1-3}
    \end{subfigure}
     \vspace{-2ex}
       \caption{Element-wise Max/Min ratio distribution of Different models (datasets).}
      \label{fig:data1}
     \vspace{-4ex}
\end{figure}

\niparagraph{Element-wise distribution.} We find that for the same element from different workers' gradient vectors at the same iteration, the relative range is also narrow. 
To demonstrate this, we analyze the distribution of element-wise $Max/Min$ ratio among eight workers' gradient vectors of the training of three models and datasets (see \secref{sec:mleva} for detailed setup and configuration), and plot the results at the early training phase (\ie, the first epoch) in \figref{fig:data1} (we have observed similar distributions through the mid/final phases of the training.).
We find that, regardless of the models and datasets, most ($\sim83\%$) elements' $Max/Min$ ratio is smaller than $2^7$.

\niparagraph{Precision loss/error tolerance.} 
It is well known that small floating point error does not dramatically affect the convergence and final accuracy of ML models~\cite{10.1109/MICRO.2018.00023,de2018high,devarakonda2017adabatch,courbariaux2014training}. 
This observation has motivated extensive prior research about training with low or mixed-precision FP operations~\cite{de2018high,micikevicius2017mixed,jia2018highly,wang2018training,NEURIPS2018_6a9aeddf,rethinkfp} and compression or quantization~\cite{10.1109/MICRO.2018.00023,horvath2019natural,10.1109/ISCA.2018.00070,HanCoRR2015}.

Thanks to these numerical characteristics, \approxarch addition can be directly applied to the in-network aggregation scenario on current Tofino switches. 
As discussed in \secref{sec:approx}, the lack of a shift-and-add operation introduces error only when adding values that differ by more than a $2^7$ ratio -- which \figref{fig:data1} shows is rare -- and the workload can tolerate such error.
We show later that it has no impact on model accuracy or convergence.
However, as discussed in \secref{sec:impl_c}, the cost of shift operations does mean the 
current Tofino only accommodates one \approxarch module per pipeline. 
Hence, in-network aggregation performance will benefit from the variable-length shift enhancement we propose.

\subsection{Evaluation}
\label{sec:mleva}

We take a two-step approach our evaluation. (1) We first show that
\approxarch addition will not affect the training convergence (\ie,
\approxarch will not incur more training iterations), and do not
consider time-wise performance. (2) We demonstrate that \approxarch
can reduce the time of each training iteration and do not consider the
convergence (because it is agnostic to per-iteration time). Taken
together, we conclude that \approxarch reduces  end-to-end training time.

\niparagraph{Benchmark.} To demonstrate \approxarch addition's impact on distributed training, we select seven popular state-of-the-art ML models.
These models are MobileNetV2~\cite{sandler2018mobilenetv2}, VGG19~\cite{simonyan2014very}, ResNet-50~\cite{he2016deep}, GoogleNet~\cite{szegedy2015going}, LSTM~\cite{jozefowicz2016exploring}, DeepLight~\cite{deng2020deeplight}, and BERT~\cite{devlin2018bert}.
We use all of these to evaluate training time, but evaluate accuracy only for the first four, since emulating \approxarch in software is costly and these CNNs train much faster than the other models.
For these CNN models, we use the CIFAR-10 dataset~\cite{cifar10} with a learning rate of 0.1, momentum of 0.9, and weight decay of 0.0005. For other models, we use the same setting as in the SwitchML evaluation~\cite{sapio2019scaling}. 
Regarding the batch size, for the accuracy experiments, we use the batch size of 16 since it leads to more frequent gradient aggregation -- if the \approxarch training convergence with 16 batch size is not affected, the convergence with larger batch size (\ie, less frequent gradient aggregation) will not be affected as well; for the performance experiments, we use the standard batch sizes of each model listed in the MLPerf benchmark~\cite{mlperf} and the SwitchML work~\cite{sapio2019scaling}.

\niparagraph{Testbed environment setup.} 
We evaluate the effectiveness of \approxarch for in-network aggregation in two ways.
To measure training accuracy and the impact of error, we write a C library that simulates gradient aggregation using a faithful implementation of the \approxarch addition algorithm and integrate this C library into PyTorch~\cite{paszke2019pytorch} to train the models.
We use the apex~\cite{apex} PyTorch extension to evaluate both FP32 and FP16 floating point formats.

To measure training time, we use an 8-machine cluster where each node is equipped with one NVIDIA P100 16 GB GPU, two 10-core Intel Xeon E5-2630v4 2.2GHz CPUs, and 128 GB of DRAM with data served from local SSD storage. The cluster is networked at 100 Gbps and includes one Tofino-based Wedge100BF-65X programmable switch. This cluster deploys in-network aggregation through SwitchML~\cite{sapio2019scaling} (due to ML framework limitations, we only evaluate FP32-based training). 
We seek to measure the performance that \approxarch can achieve with our variable-length shift extension, which allows multiple parallel \approxarch instances per pipeline. Because current switch hardware does not support this, we emulate \approxarch-enabled performance by removing the end-host format conversion/quantization at the workers and performing integer computations in place of FP computations on the switch.
While this emulation setup gives nonsensical output, it provides a realistic expectation of \approxarch performance because: (1) under Tofino, data plane programs experience a switch processing latency that depends only on the number of stages and not on the computation intensity of their specific operations, without any effect on throughput (data plane programs operate at line rate) as confirmed experimentally in previous work (e.g.,~\cite{Dang.P4xos});
(2) SwitchML uses the full set of stages on the ingress pipelines of Tofino and any potential increase of in-switch latency can be mitigated by increasing the number of aggregation slots.
Note that we use this approach only for performance evaluation.

\begin{figure}[t]
  \begin{subfigure}[b]{0.32\linewidth}
    \includegraphics[width=\textwidth]{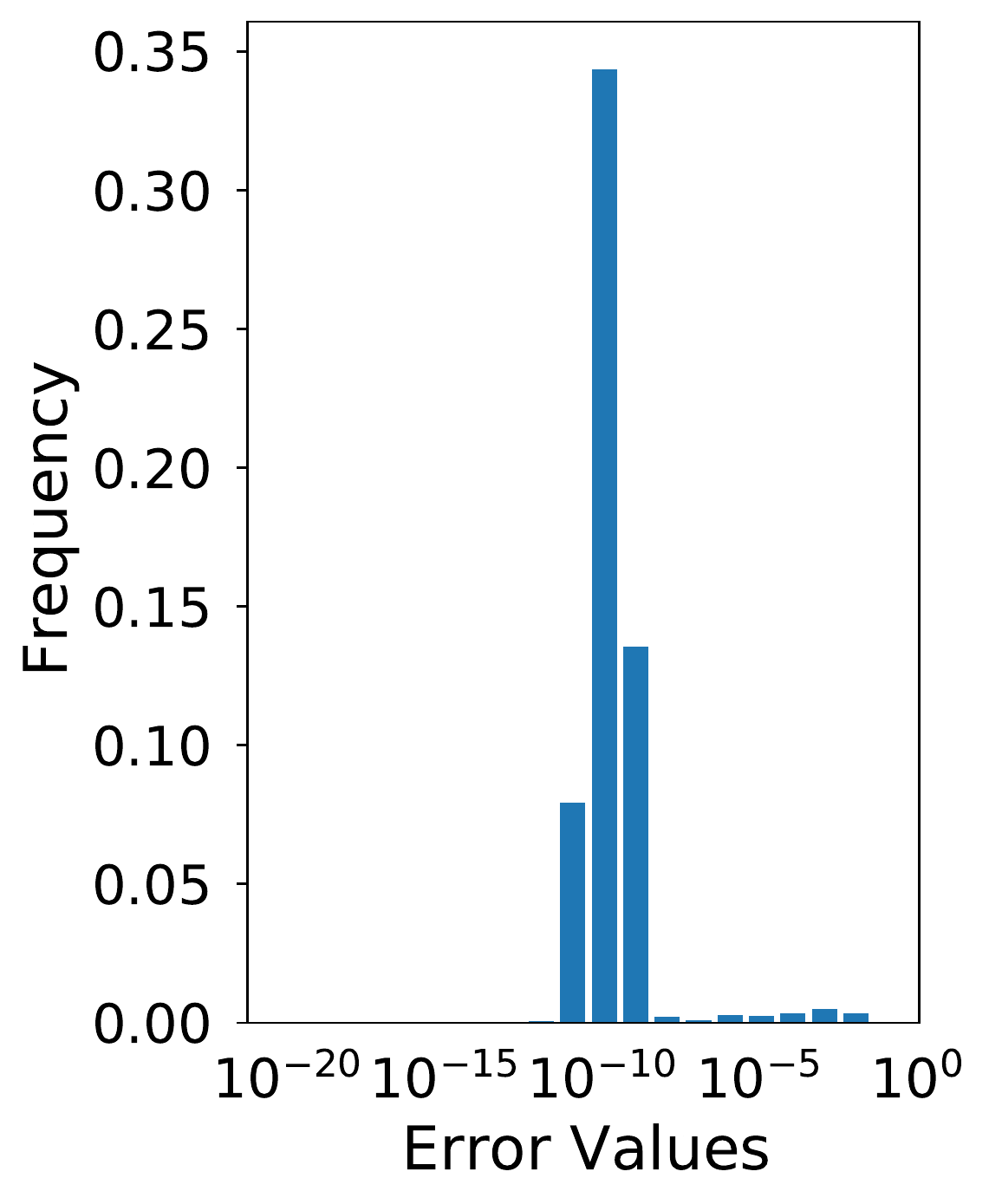}
    \caption{Epoch \#1.} 
    \label{fig:data2-1}
  \end{subfigure}
   \hfill
  \begin{subfigure}[b]{0.32\linewidth}
    \includegraphics[width=\textwidth]{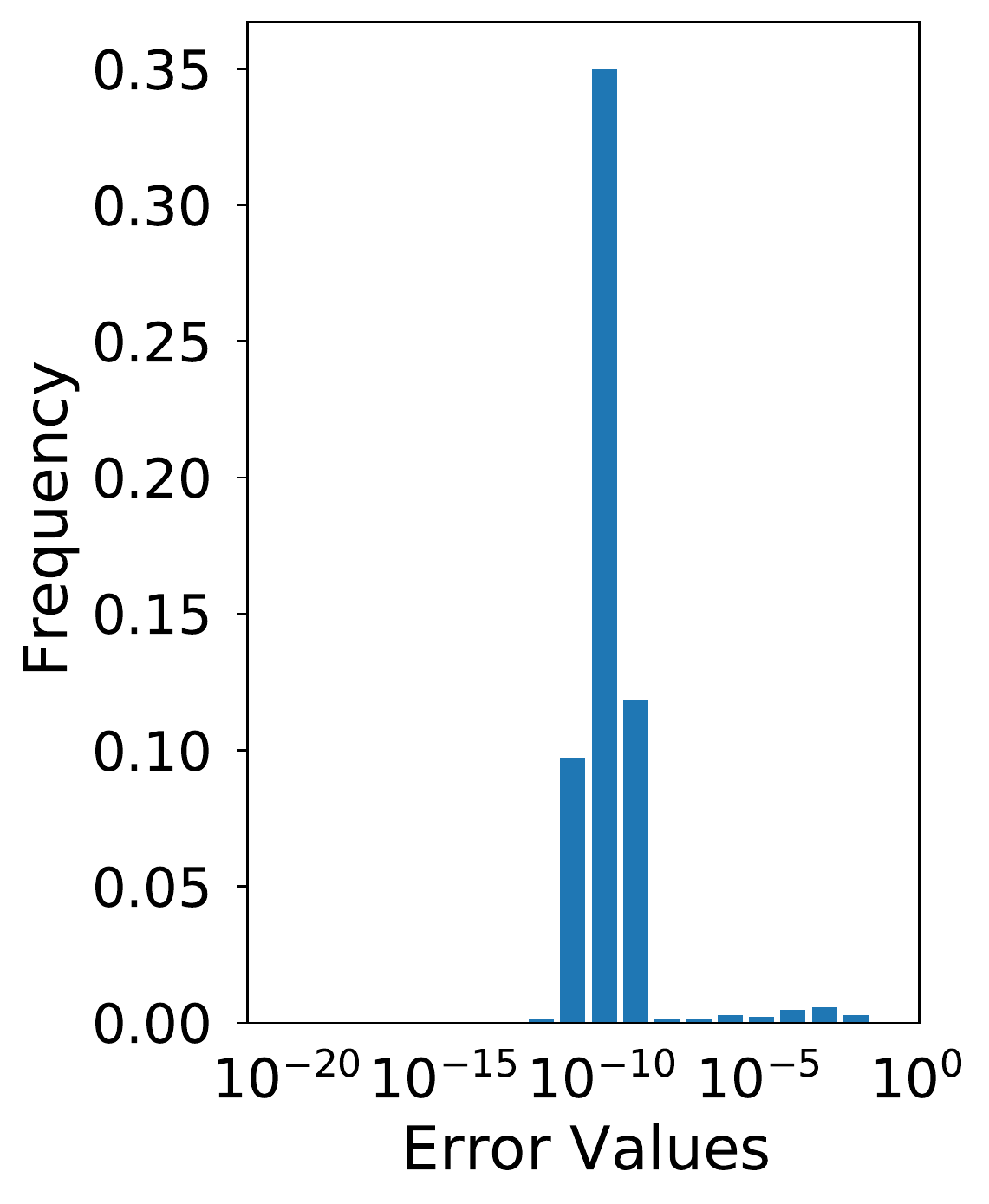}
      \caption{Epoch \#20.}
    \label{fig:data2-2}
  \end{subfigure}
   \hfill
  \begin{subfigure}[b]{0.32\linewidth}
    \includegraphics[width=\textwidth]{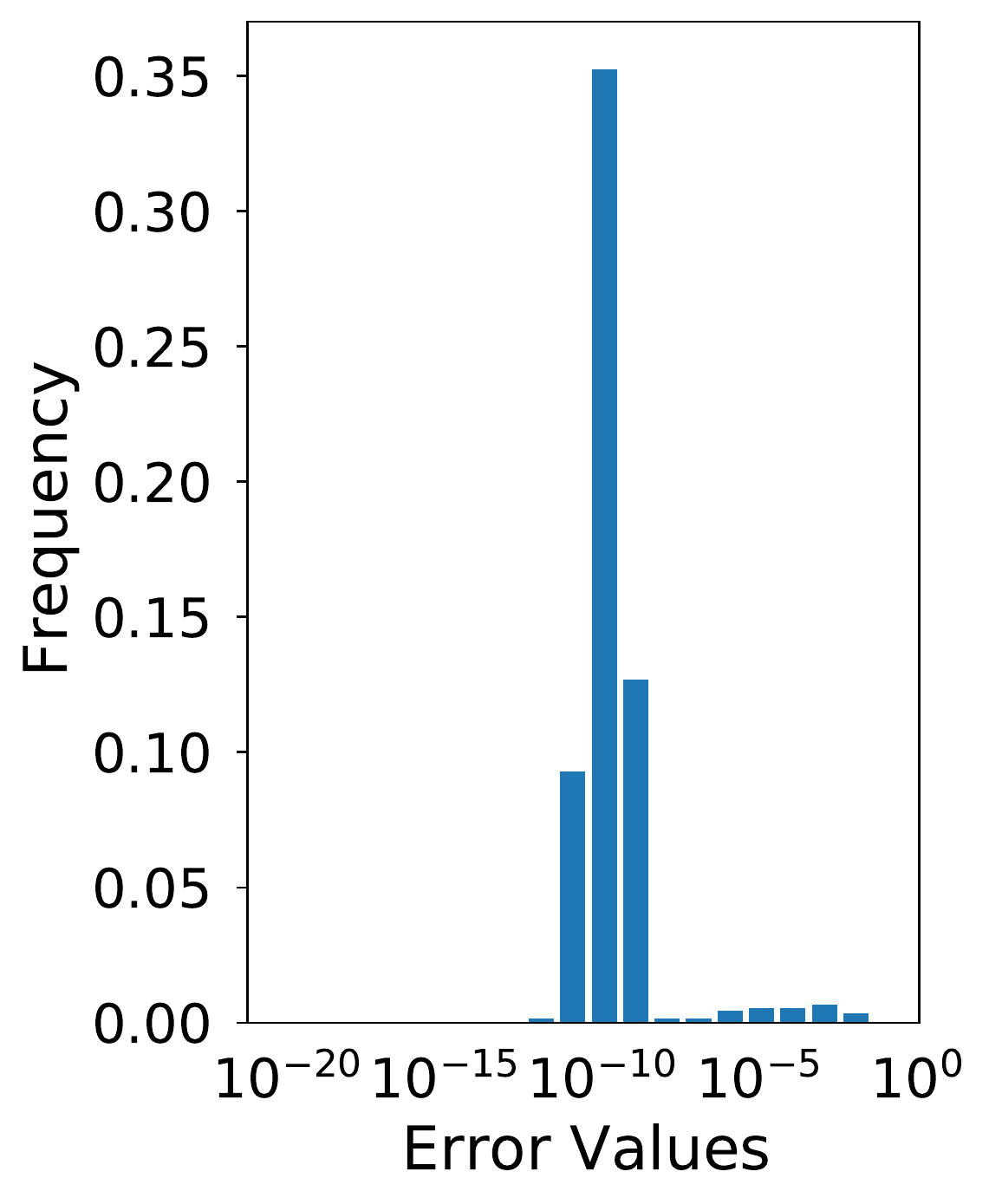}
    \caption{Epoch \#40.}
    \label{fig:data2-3}
  \end{subfigure}
   \vspace{-2ex}
   \caption{\approxarch's error distribution of VGG19 gradient aggregation at early, middle, and final training stages.}
    \label{fig:data2}
   \vspace{-4ex}
\end{figure} 
\subsubsection{\approxarch Error Analysis}
\label{sec:error}
To investigate the errors to which \approxarch addition may lead, we record the gradient vectors from eight workers during a training job. 
We use the software \approxarch library to compare the results of  \approxarch vs. standard floating point addition for aggregating the same gradient vectors. 
\figref{fig:data2} shows the (absolute) error distribution of VGG19 during different training phases.

Similar to the gradient distribution~\cite{10.1109/MICRO.2018.00023}, the error distribution remains similar among early, middle, and final phases of  training, showing \approxarch's wide applicability.
Most errors (>95\%) are in the range of $\text{[}10^{-10},10^{-8}\text{]}$, enough to be tolerated by ML training, which we demonstrate in the next section. 
We further investigate the sources of the errors and find that most errors come from rounding, while the errors caused by the overwrite and left-shift mechanisms happen rarely (less than 0.9\% and 0.1\%, respectively, among all the addition operations in the aggregation procedure).
These errors arise because, in some cases, a gradient vector's element-wise distribution is larger than \approxarch's left-shift headroom. 
As a result, the smaller values may be ignored in the aggregation procedure, leading to small errors (\ie, smaller than $10^{-8}$).

Note that a switch implementing the full \arch proposal, rather than just \approxarch, would not experience these overwrite errors. Note also that no overflow occurs in this experiment, since the number of workers, and thus the number of operations per vector element, is less than the headroom available in the mantissa register.

\begin{figure*}[t]

    \begin{subfigure}[b]{0.24\textwidth}
      \includegraphics[width=\textwidth]{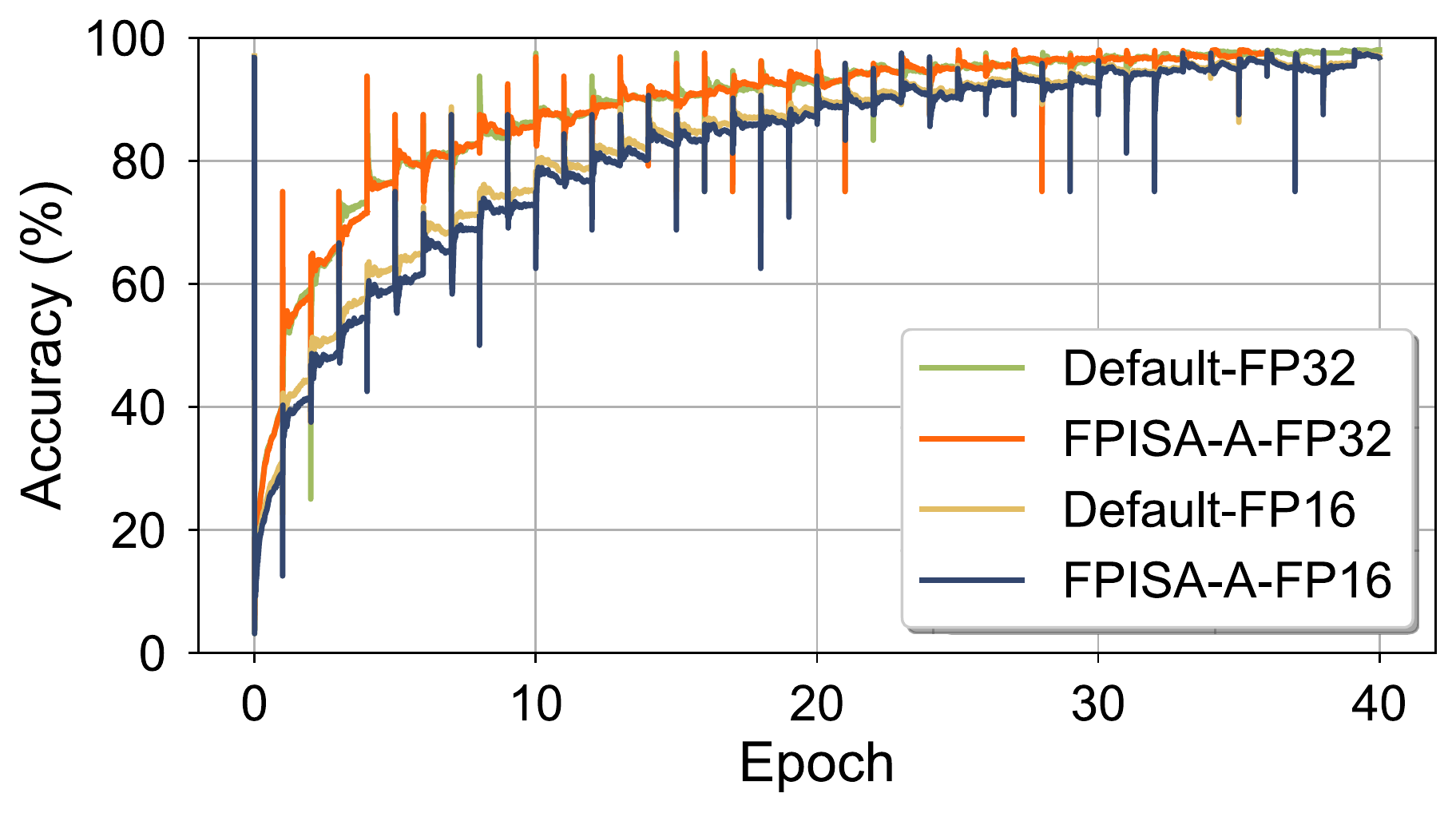}
      \caption{GoogleNet.} 
      \label{fig:loss-google}
    \end{subfigure}
     \hfill
    \begin{subfigure}[b]{0.24\textwidth}
      \includegraphics[width=\textwidth]{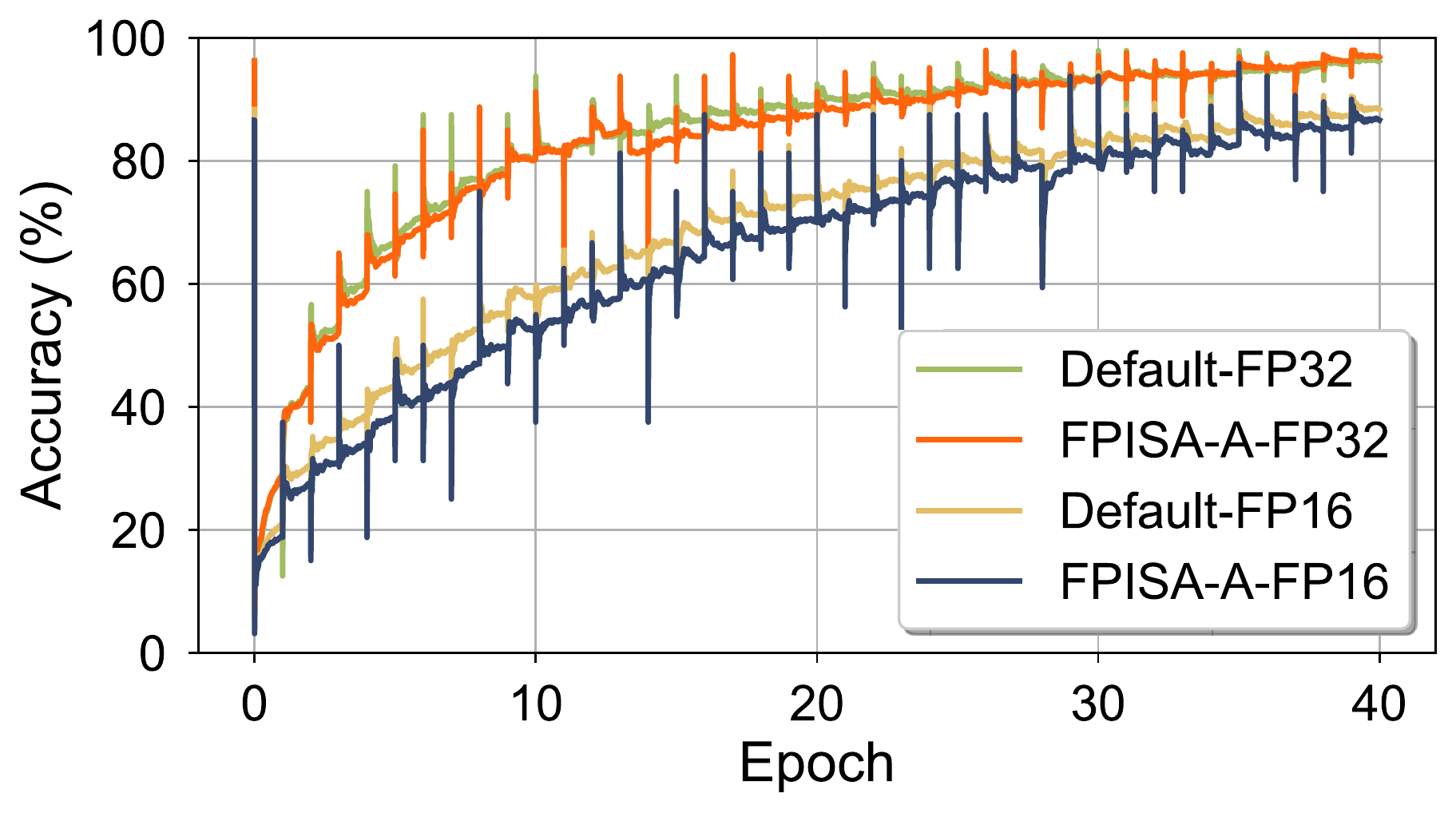}
        \caption{ResNet-50.}
      \label{fig:loss-res}
    \end{subfigure}
     \hfill
    \begin{subfigure}[b]{0.24\textwidth}
      \includegraphics[width=\textwidth]{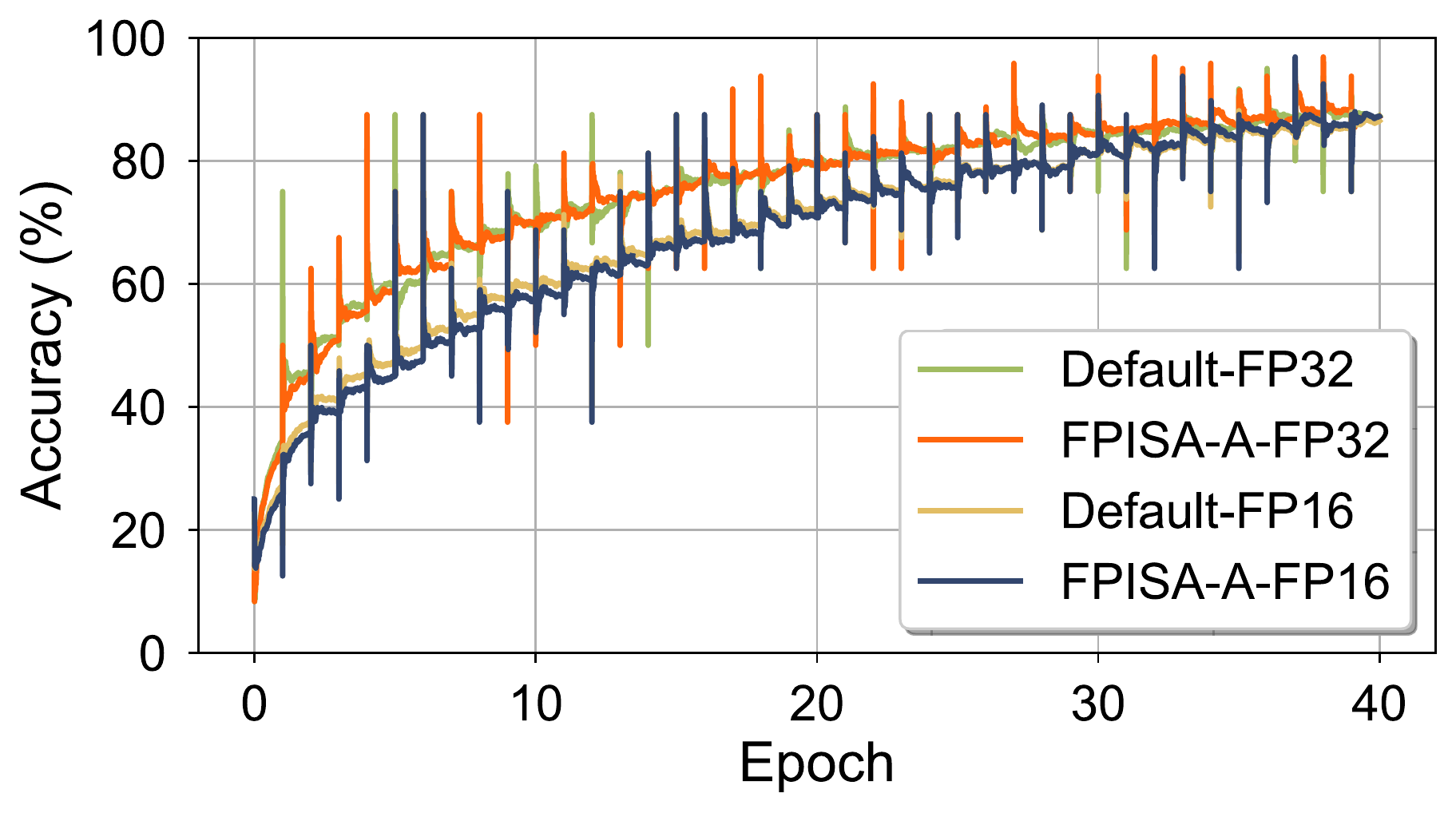}
      \caption{VGG19.}
      \label{fig:loss-vgg}
    \end{subfigure}
    \hfill
    \begin{subfigure}[b]{0.24\textwidth}
        \includegraphics[width=\textwidth]{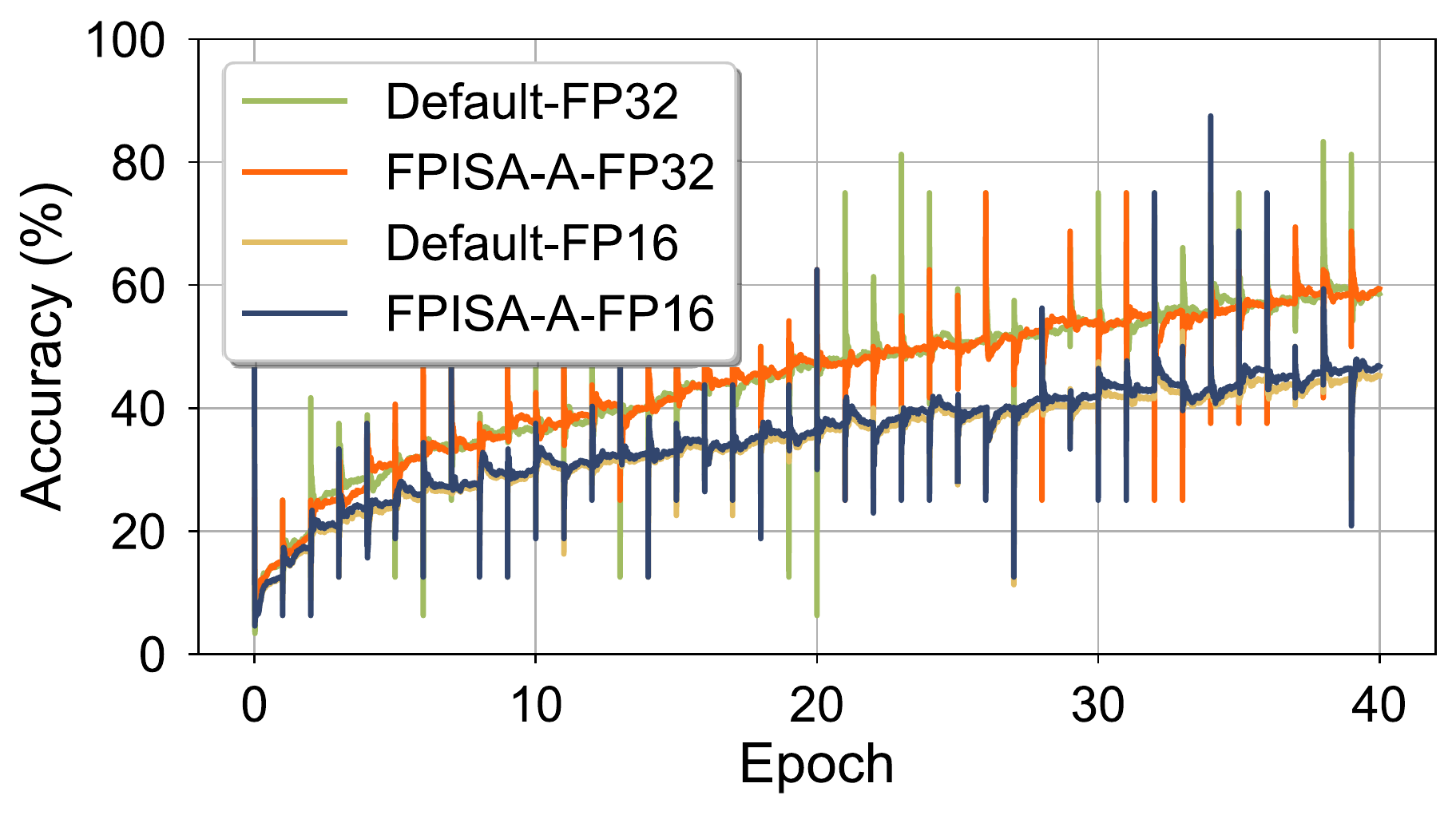}
        \caption{MobileNetV2.}
        \label{fig:loss-mobilenetv2}
      \end{subfigure}
     \caption{Accuracy curves of different ML models with default addition and \approxarch addition.}
      \label{fig:all-acc}
\end{figure*}

\subsubsection{\approxarch's Impact on Training Convergence}
We investigate whether \approxarch will lead to training accuracy loss, due to the errors it imposes.
We train four ML models for 40 epochs with both default and \approxarch addition in gradient aggregation. 
To show \approxarch's adaptability on different floating point formats, we train using both standard single-precision FP32 and half-precision FP16 for each model.

We plot the accuracy value during the training procedures of each model in \figref{fig:all-acc} to observe \approxarch's impact on convergence.
Note that the jitters in the curves are due to the small batch size we are choosing; these are normal and do not affect the training procedure.
First, we find that floating point precision does affect the training convergence. 
That is, in all four models, we observe slower convergence of FP16-based training compared to regular FP32-based training, as well as the final accuracy. 
However, \approxarch's addition errors will not amplify such gaps. 
In most cases, the curve of \approxarch addition is closely aligned with the curve of default addition.
After 40 epochs, the accuracy differs by less that 0.1\% 
The results also demonstrate that regardless of the floating point format, \approxarch addition will not degrade each model's accuracy.
Hence, we argue that \approxarch will not prolong the training by adding necessary epochs to convergence.

\begin{figure}[!t]
  \centering 
  \includegraphics[width=\linewidth]{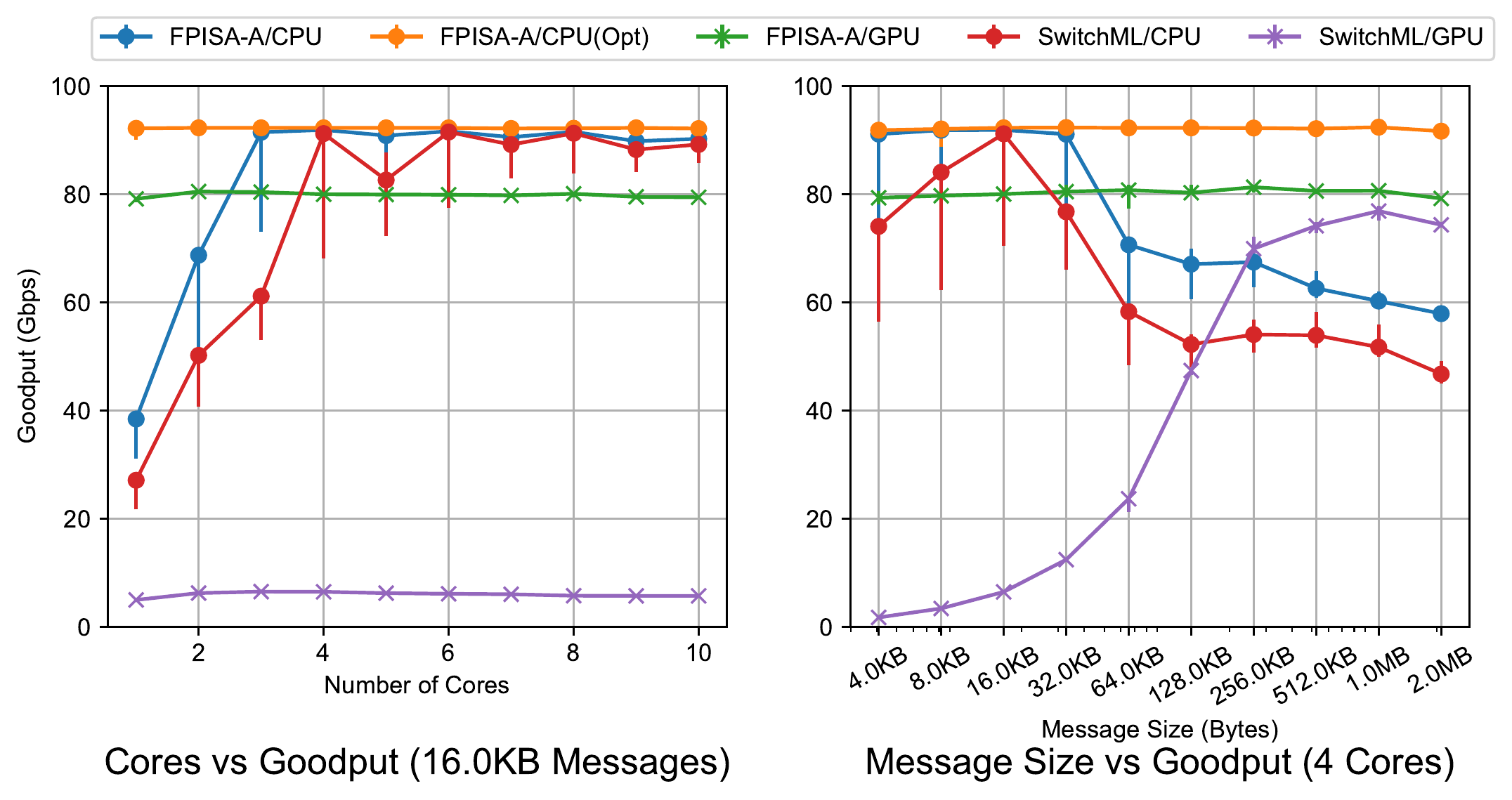}
  \caption{Goodput of different FP approaches on microbenchmark. The maximum theoretical goodput with framing overhead is 92 Gbps.}
  \label{fig:micro-bench}
\end{figure}

\begin{figure}[!t]
  \centering 
  \includegraphics[width=\linewidth]{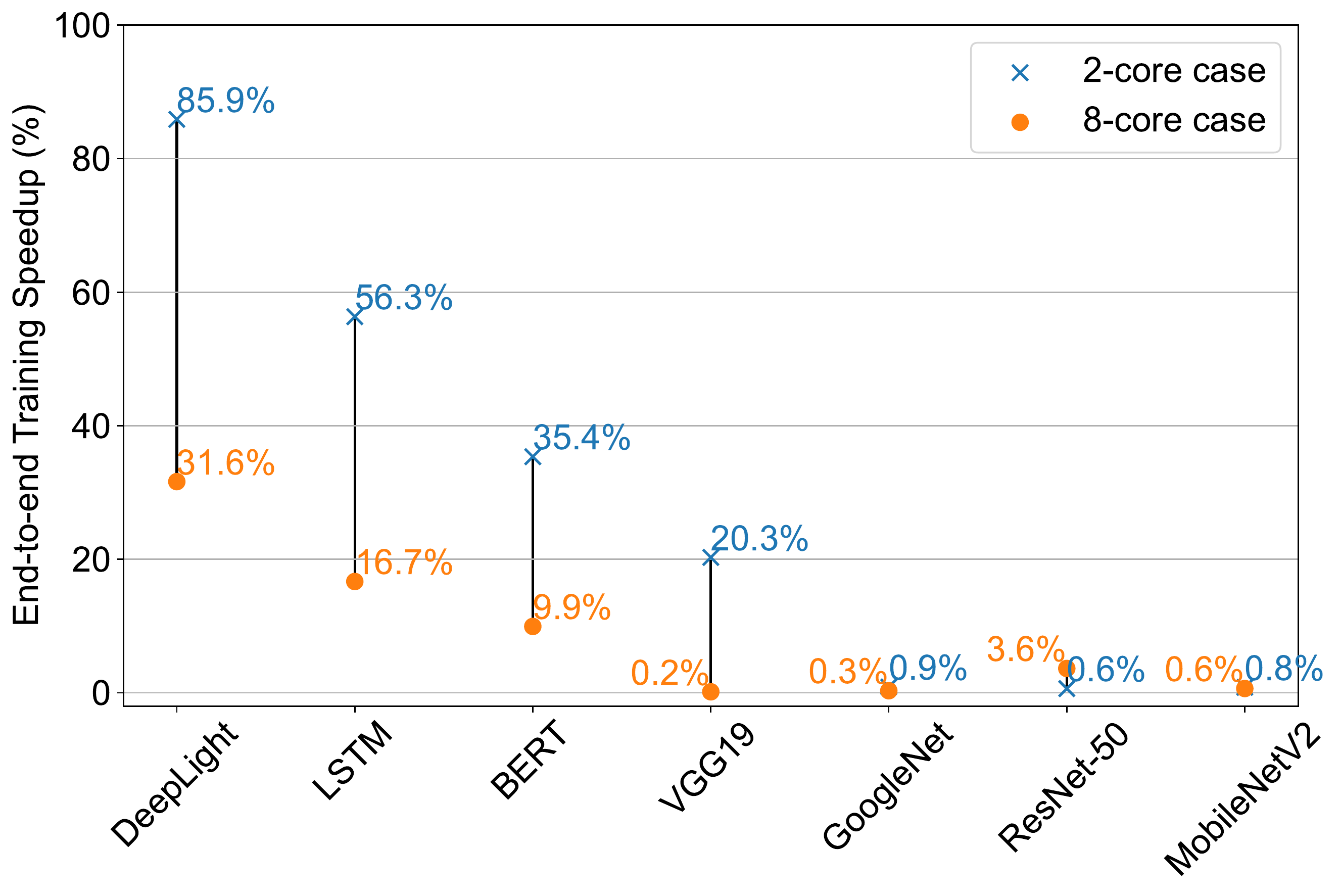}
  \caption{End-to-end training time speedup of \approxarch compared to the default SwitchML.}
  \label{fig:e2e}
\end{figure}

\subsubsection{Training Speedup with \approxarch}
\label{sec:mlspeedup}

In the next experiments, we evaluate the potential speedup of \approxarch in an end-to-end training setting as well as the resulting reduction of host-based quantization overheads.
SwitchML uses CPU cores at workers to scale and transform the numeric representation of gradient vectors, including both floating-point/integer conversion and byte order conversion. In contrast, \approxarch does not have these overheads as it sends gradient vectors as floating point values directly. Thus, we vary the number of CPU cores and measure the throughput differences between these approaches through a microbenchmark.

In this microbenchmark, two workers reduce a 1GB gradient vector;\footnote{We use two workers to exclude the synchronization variability among a larger number of workers. This is to better quantify the performance differences due to the scaling and transformation overheads. We also tried 100MB with similar results.} we measure the time to complete the operation across the workers. We use 256 element packets which is the largest that SwitchML supports. After 50 warm-up invocations, we perform 250 reductions and report median and 90th-10th percentiles as the error bars.

We use SwitchML's RDMA transport since it is more efficient that the DPDK one, and run two versions to explore the performance implications of scaling and transforming gradient vectors on either the CPU or the GPU (where gradients are initially computed and stored).
The base SwitchML version -- denoted SwitchML/CPU -- uses CPU cores. This benchmark assumes that the gradient vectors are already in host memory.
Further, we create a new version of SwitchML -- denoted SwitchML/GPU -- that uses the GPU to scale and transform gradient vectors to the integer representation before copying them to pinned host memory. Recall that SwitchML scales the gradient vectors in chunks, using a scaling factor adapted to each chunk based on a maximum exponent calculation 
that overlaps with the aggregation of the previous chunk. For SwitchML/CPU, we keep the original SwitchML logic where one chunk is equivalent to the RDMA message size. For SwitchML/GPU, we use a separate CUDA stream for each core used to allow parallel kernel execution. We also introduce a performance optimization where we asynchronously dequantize aggregated messages from integer into floating point values on a separate CUDA stream thus having 2 CUDA streams for each core used.

We run our \approxarch emulation in three settings. (1) \approxarch/CPU adopts the RDMA implementation of SwitchML and disables host-based type conversions. SwitchML's RDMA implementation, however, involves a memory copy operation into a staging area. Given that it should be possible for \approxarch to operate entirely on memory-resident native FP vectors, we include a further optimization -- (2) \approxarch/CPU(Opt) -- that emulates the case where the memory copy is not necessary.
Lastly, (3) \approxarch/GPU includes a copy from GPU memory to pinned host memory and back.\footnote{Our testbed does not support GPU Direct, which would enable \approxarch to use RDMA transfers out of and into GPU memory.}
We introduce three performance optimizations for this approach: we use batching to amortize the cost of launching one copy operation for each chunk, we asynchronously kick off copies from GPU to host memory always one batch ahead of what needs to be consumed, and we asynchronously copy back from host to GPU memory on a separate CUDA stream similar to the SwitchML/GPU case. Note that the batching and copying one batch ahead optimizations cannot be trivially applied to SwitchML/GPU since it needs to wait for exponents to come back for each chunk before performing any quantization/dequantization. In contrast, they can be applied for \approxarch/CPU by using DMA engines that perform copies in the background with complete overlap with CPU and network operation, further improving \approxarch/CPU performance to possibly match \approxarch/CPU(Opt).

Starting with the CPU variants, \figref{fig:micro-bench} (left) shows that \approxarch/CPU requires only three CPU cores to achieve the 92 Gbps maximum goodput, as opposed to SwitchML/CPU, which needs four cores (\ie, \approxarch uses $\sim$25\% fewer cores than SwitchML).\footnote{SwitchML/CPU with 5 cores has a small performance dip due to work imbalance across cores in this particular configuration.}
\approxarch/CPU(Opt) achieves the maximum goodput with just a single core ($\sim$75\% fewer cores than of SwitchML).
This leaves more CPU cycles for data I/O, potentially avoiding training job stalls while waiting for input data to be preprocessed.

For the GPU variants, \figref{fig:micro-bench} shows that SwitchML/GPU is inefficient with message sizes below 256KB. This is due to overheads of GPU kernel launches and copies at small message sizes. Increasing the number of cores does not help because CUDA implicitly synchronizes all kernel launch calls (kernel execution can be parallelized whereas kernel launches cannot). In contrast, using just a single CPU core, \approxarch/GPU can achieve the best possible performance -- limited to 80 Gbps only by the bidirectional copy bandwidth of the GPU copy engines  -- since it can copy messages in larger batches.\footnote{We copy memory using 1MB chunks as it gives the best results irrespective of the RDMA message size.} We expect that without this bidirectional copy bandwidth limit (a constraint of our GPUs for the copy sizes in use), \approxarch/GPU would match the performance of \approxarch/CPU(Opt) since it completely overlaps the memory copying with CPU and network operations.

\figref{fig:micro-bench} (right) shows that SwitchML/GPU with a copy size of 1MB reaches a performance comparable (but still below) to \approxarch/GPU. However, this requires an equally large RDMA message size whereas \approxarch/GPU performs well even with 4KB messages.
Using large message sizes has several negative implications. First, it can introduce larger errors in SwitchML's quantization scheme since it chooses the scaling factor from a larger chunk. Second, it hurts the performance of loss recovery because the loss of a single packet entails resending the entire 1MB message (1024 packets). Third, the performance degrades past a certain message size. This is due to limited network capacity and the reduction of pipelining, which in turn reduces the performance benefits of SwitchML's streaming aggregation. Thus, we conclude that, although performing quantization on the GPU might still be an interesting possibility for SwitchML, more work is necessary to devise an efficient implementation without increasing quantization errors and without affecting the GPU's availability for training.

We now confirm that \approxarch's benefits translate into higher end-to-end training throughput.
\figref{fig:e2e} reports the training throughput for seven real-world DNN benchmarks. For these experiments, we restrict the comparison to the DPDK implementation because SwitchML/RDMA is not currently integrated into the ML frameworks~\cite{sapio2019scaling}.
We focus on two scenarios -- using either two or eight cores -- and we measure the speedup in terms of training throughput (samples/s).
We observe that \approxarch speeds up training by up to 85.9\% and 31.6\% for the 2-core case and the 8-core case, respectively.
Importantly, the higher speedup factors are obtained when using just two cores for communication, which frees up six cores for data I/O in this setting.
The speedup is particularly significant in communication-bottlenecked models (\eg, DeepLight, LSTM, BERT, VGG19), where \approxarch is up to 85.9\% faster compared to SwitchML when using the same number of cores.
On the other hand, we do not see significant benefits of \approxarch on models like GoogleNet, MobileNetV2, and ResNet-50, which are compute-bottlenecked.

By combining the accuracy results and the per-iteration end-to-end results, we can conclude that \approxarch is able to reduce the total end-to-end training time of a wide range of ML models.

%% file: db.tex
\section{Case Study: Distributed Database Queries}
\label{sec:db}

\begin{figure}[!t]
  \centering 
  \includegraphics[width=\linewidth]{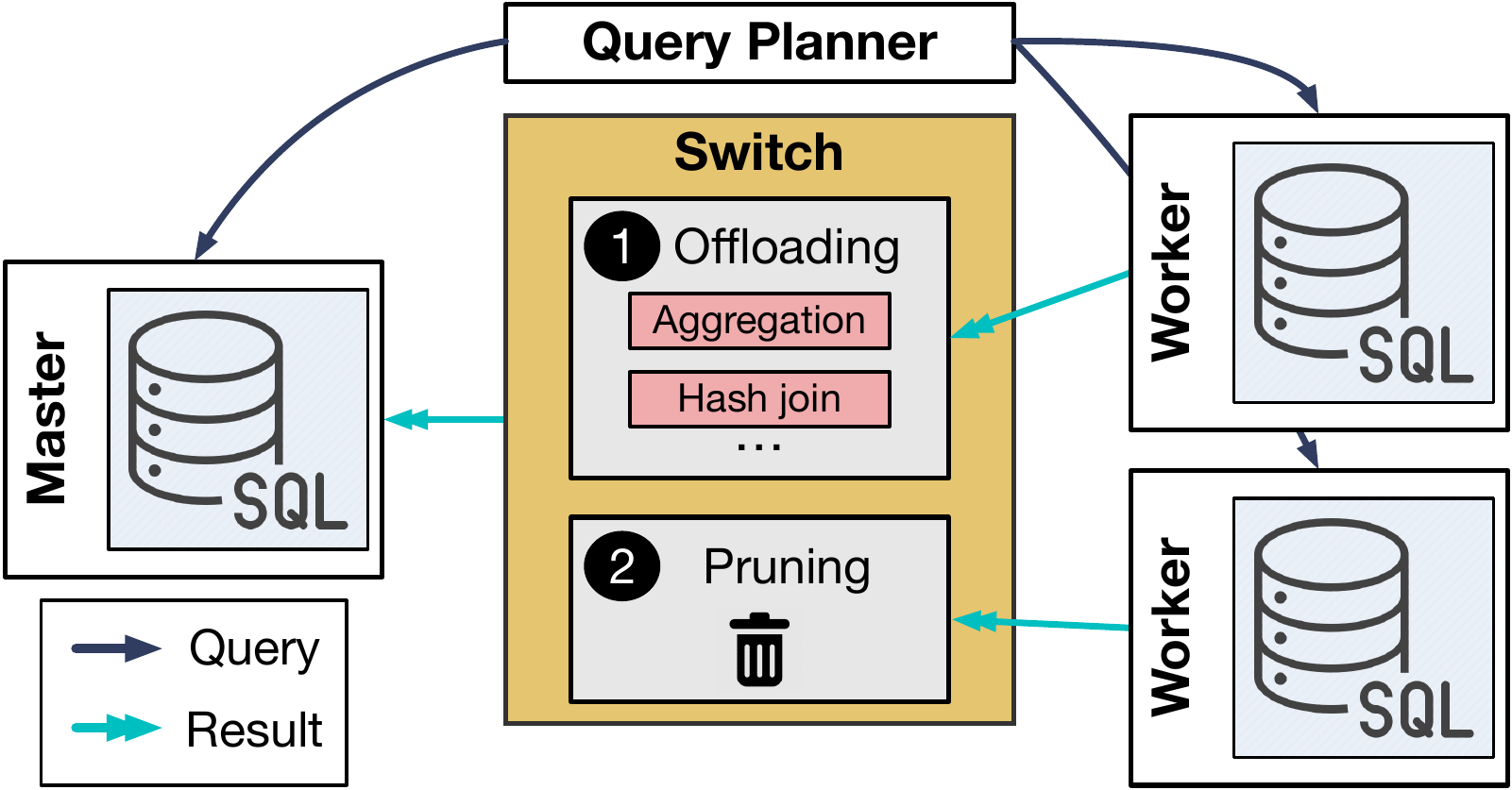}
  \caption{In-network acceleration for distributed queries.}
  \label{fig:db}
\end{figure}

Growing data volumes have created great demand for distributed databases and similar distributed data processing systems. 
In-network query acceleration can greatly improve performance for these databases.

We depict a typical distributed database system in \figref{fig:db}. 
A query planner parses and plan a query across multiple worker machines.
Each worker operates on its own data partition, and workers' results are aggregated to a master machine to form the final result. 
Server processing capabilities, along with network congestion at the
master machine, can become bottlenecks~\cite{10.1145/2723372.2742797}.

There have been two lines of research leveraging PISA switches to accelerate such distributed database queries. Neither support floating point datatypes. 
The first, NETACCEL~\cite{lerner2019case}, offloads parts of the query (\ie, query operators) to the switch (\circled{1} in \figref{fig:db}).
For example, a switch can be responsible for \textsc{join}ing rows from different workers and \textsc{group}ing the data to final results. 
In this approach, caching, hashing, and addition are the most common in-switch operations. 

The second, Cheetah~\cite{10.1145/3318464.3389698}, uses ``pruning,'' a new abstraction that drops a large portion of unnecessary data at the switch, letting the remaining data pass to the master machine for regular processing (\circled{2} in \figref{fig:db}).
For instance, in a ``Skyline'' query, the switch will cache initial data and compare it with later data.
If the later data is ``behind'' (\ie, smaller in certain dimensions) the cached data, it will be dropped. 
In this approach, caching and comparison are the major in-switch operations.

Both approaches lack floating point support, which is common in databases. 
For example, Tableau reports that 20\% of data in their hosted business
intelligence databases are floating point values, nearly as
common as integers (25\%)~\cite{vogelsgesang18:_get_real}.

\subsection{Characteristics of Query Data}
Query data processing in the switch exhibits very different characteristics from the gradient-related operations in distributed training.
First, it requires accurate results rather than tolerating approximation. 
Second, only a few elements will appear in a single packet for processing (\eg, one for Top-N and two or more for Join or Group-by~\cite{10.1145/3318464.3389698}), demanding little parallelism inside the switch.
Finally, there is no characteristic numerical distribution of the values processed by the switch (since the data can be arbitrary), unlike distributed training.

Due to its accuracy requirements, this application domain is not
suitable for the \approxarch approximation approach. It requires,
instead, the full \arch approach, including hardware modifications --
our proposed shift+add ALU operation -- to provide sufficient accuracy
for this application.

\begin{table}[!tb]
  \centering
  \caption{Description of evaluated queries.}
  \vspace{-1ex}
\scriptsize
  \label{tab:query}
  \begin{tabular}{m{3.5cm}cc}
    \toprule
   \bf Query &  \bf Acceleration method & \bf FP operation\\
  \midrule
  Top-N & In-switch pruning & Comparison\\
  Group-by-having max/min & In-switch pruning & Comparison\\
  Group-by (hash-based aggregation) & In-switch aggregation & Addition\\
  TPC-H Q3 & In-switch pruning & Comparison\\
  TPC-H Q20 & In-switch aggregation & Addition\\
  \bottomrule
  \end{tabular}
  \vspace{-3ex}
\end{table}
\subsection{Evaluation}
\niparagraph{Benchmarks.} We select five queries from Cheetah~\cite{10.1145/3318464.3389698} and NETACCEL~\cite{lerner2019case} (with the same settings) and modify them to use FP32 datatypes. \tabref{tab:query} describes the details. We include only the queries that are affected by in-switch floating point operations; for other queries, our performance would be equivalent to the original work. 
For single-operation queries, we use the dataset from Big Data~\cite{10.1145/1559845.1559865}, which contains 30M rows in the uservisits table and 18M rows in the rankings table. The queries themselves are the same as described in Cheetah~\cite[Appendix B]{10.1145/3318464.3389698}.
We also test Query 3 and 20 in TPC-H benchmark~\cite{tpch} (default scale factor), the two queries evaluated in Cheetah~\cite{10.1145/3318464.3389698} and NETACCEL~\cite{lerner2019case}.
We convert the corresponding datatype in both Big Data (the column ``adRevenue'' in the uservisits table) and TPC-H (the column ``L\_extendedprice'') from \texttt{int32} to FP32. 

\niparagraph{Testbed environment setup.} We extend Cheetah's DPDK-based end-host program and compare its performance against the Spark-based baseline. 
We deploy the program and Spark to one master machine and two worker machines, all connected to the Tofino switch via Intel X710 40GbE NICs. 
Each worker machine stores the database partition in a \texttt{tmpfs}
RAM disk, and we ignore the first run of Spark since it is unstable~\cite{188988,10.1145/2723372.2742797,10.1145/3318464.3389698}.

\begin{figure}[!t]
  \centering 
  \includegraphics[width=\linewidth]{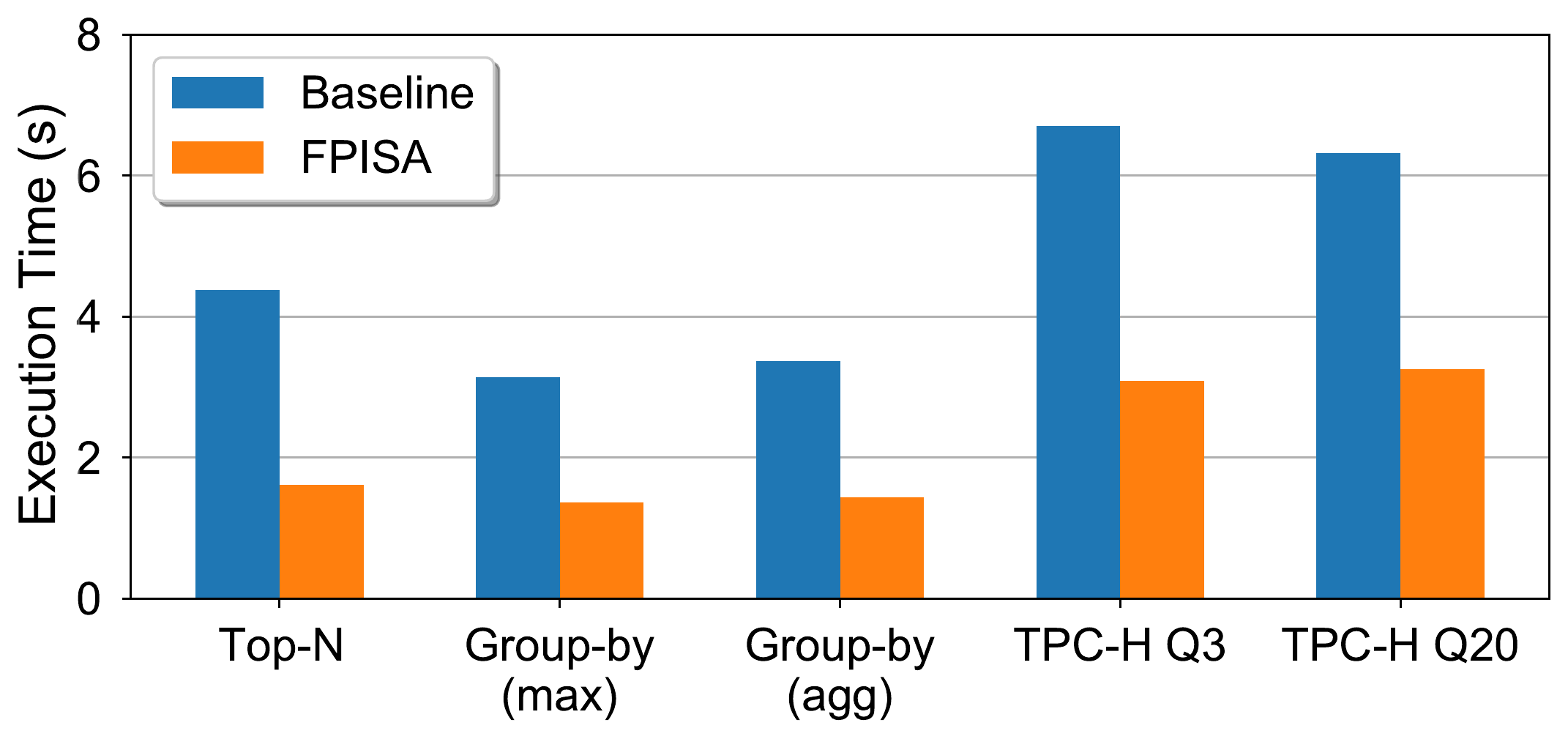}
  \caption{Performance of distributed DB queries with \arch.}
  \label{fig:db-res}
\end{figure}

Due to the limitations of current hardware described earlier, we 
must use an emulation approach to evaluate \arch's performance on
database acceleration. We first run the Cheetah system using integer
data types to record which packets are pruned or aggregated in the
switch. We then run the Cheetah benchmark itself, using FP32 data
types. We tag the packets sent with additional header bits to indicate
the recorded action. Our switch uses these to take the correct action
that a \arch switch with our improvements would perform. During this
benchmark, we measure the end-to-end execution time on the master
machine.

This emulation approach gives a faithful measurement of end-to-end
performance. It does not change
(1) the size of data transferred in the network; (2) the performance
of the programmable switch, as its task complexity does not affect
processing performance, and (3) the processing costs at the end hosts,
which run the existing Cheetah code.

\niparagraph{Results.}
We present the speedup of distributed queries using floating point datatypes in \figref{fig:db-res}.
The results are highly aligned with their integer-based counterparts~\cite[Fig. 8]{lerner2019case}~\cite[Fig. 5]{10.1145/3318464.3389698}, yielding a speedup of 1.9--2.7$\times$ depending on the query over a Spark baseline. (Spark's local compute time for integer and floating point datasets differs by less than $\pm 3\%$.) 
This means with \arch's help, distributed queries with floating point datatypes can be accelerated by switches just as with integer datatypes.

%% file: related.tex
\section{Related Work}
\label{sec:related}

\niparagraph{Accelerating distributed applications with programmable switches.}
Recently, programmable switches have been used to accelerate a broad range of applications, including distributed key-value stores~\cite{10.1145/3132747.3132764,tokusashi18:_lake,li:_pegasus}, distributed transactions~\cite{10.1145/3132747.3132751,211261,10.1145/3387514.3405857}, distributed storage~\cite{10.14778/3368289.3368301,227794}, packet queuing/scheduling~\cite{211285,10.1145/2934872.2934899}, network functions~\cite{10.1145/3387514.3405855,10.1145/3098822.3098824}, and network telemetry~\cite{10.1145/3190508.3190558,10.1145/3387514.3405894,10.1145/3387514.3406214,10.1145/3230543.3230555}.
While most of them deal with packet header processing with few arithmetic operations, some perform computation on the packet's payload.
SwitchML~\cite{sapio2019scaling} and ATP~\cite{265053} leverage switches for gradient aggregation but are constrained to fixed-point aggregation, which may lead to costly format conversion on the end-host and additional network round trips for exponent communication. 
NETACCEL~\cite{lerner2019case} and Cheetah~\cite{10.1145/3318464.3389698} propose to use programmable switches to accelerate database queries by data pruning or query offloading, but FP datatypes are not supported.

\niparagraph{Resource allocation.} Much research has studied how to
use in-network rate computations to support congestion control (e.g., XCP~\cite{xcp} and RCP~\cite{rcp}), queue management (e.g., CoDel~\cite{codel} and AIFO~\cite{10.1145/3452296.3472887}), or load balancing (e.g., CONGA~\cite{conga}). P4QCN~\cite{p4qcn}, P4-CoDel~\cite{p4codel}, and P4-ABC~\cite{p4abc} are P4 implementations of specific protocols that require floating point support -- currently unavailable in switch hardware. Sharma \etal proposed a library that applies approximation to work around this limitation~\cite{201474}.
InREC~\cite{joseinrec} and NetFC~\cite{cui2021netfc} proposed to use table-lookup for floating point operation emulation in programmable switches. However, they are constrained to stateless operations and need extra RAM space to store the tables. Also, few floating point operations can be done per packet, limiting  parallelism. \arch may enable new design options for in-switch resource allocation.

\niparagraph{Extending switches' processing capability.}
Proposed enhancements to the RMT architecture~\cite{10.1145/2486001.2486011} include transactions~\cite{10.1145/2934872.2934900}, disaggregated memory~\cite{10.1145/3098822.3098823}, and better stateful data plane support~\cite{10.1145/3422604.3425928}.
While many focus on improving stateful computations, none address floating point operations.

\niparagraph{Advanced floating-point operations.}
This paper describes addition and comparison, two commonly-used
floating point operations. Other more complex floating point
operations may be needed for future applications (\eg, congestion control~\cite{xcp,rcp} and network security~\cite{10.1145/3230543.3230555}). 
Appendix~\ref{sec:advancedops} briefly discusses the possibility of supporting them.

%% file: conclusion.tex
\section{Conclusion}
\label{sec:concl}

In this work, we propose \arch, a floating point representation designed to work efficiently in programmable switches. 
We first implement \arch on a commodity Intel Tofino switch, but its design limits throughput and accuracy.
We then propose hardware changes based on the Banzai programmable switch architecture to avoid these limitations.
We demonstrate their feasibility through synthesis using a 15-nm standard-cell library, and find minimal impact on area, power, and timing.
Finally, we investigate the benefit of \arch by implementing accelerators for two applications, evaluating their performance on a switch implementing our changes using emulation. We find that \arch allows distributed training to use 25-75\% fewer CPU cores and provide up to 85.9\% better throughput in a CPU-constrained environment than the state-of-the-art framework. For distributed query processing with floating point data, \arch enables up to 2.7$\times$ better throughput than native Spark.

%% file: appendix.tex
\appendix

\section{Additional Floating Point Features and  Operations}
\label{sec:advancedops}

\subsection{Floating point semantics}

\niparagraph{Rounding.} For simplicity, we have described \arch without guard digits. The combination of no guard digits and two's-complement representation provide round-toward-negative-infinity semantics. An implementation with $n$ guard digits would simply store the mantissa shifted left $n$ bits from what is show in~\figref{fig:format}, and would use those to perform other types of rounding after renormalization. 

\niparagraph{Reproducibility.} \arch provides reproducibility in that the same sequence of operations and values will always produce the same result. However, since \arch performs operations in a different order than that specified in the IEEE 754 standard, the same sequence of operations and values performed on an IEEE-754-compliant CPU may yield a different result than \arch. For the use cases we describe in this paper, IEEE 754 compliance is not a requirement.

\subsection{Advanced FP operations}

In this paper, we have covered the two commonly-used floating point operations -- addition and comparison.
They are sufficient for many distributed applications.
However, other more complex and costly floating point operations may be needed in the future with emerging applications (\eg, congestion control~\cite{xcp,rcp} and network security~\cite{10.1145/3230543.3230555}). 
To pave the way for future PISA implementations, we briefly discuss the possibility of supporting them.

\niparagraph{Multiplication and division.} 
The flow of floating point multiplication is similar to that of addition in \secref{sec:add}.
The two major differences are (1) the two exponents are added, and (2) the two mantissas are multiplied, all as integers.
For small floating point types, the mantissa multiplication can be implemented as a table lookup, without hardware modifications.
For larger floating point types, integer multiplers could be added to the hardware. We implement one based on Banzai and find its overhead is acceptable: approximately the same as an adder and a boolean module \wrt power and area. 

Floating point division has a different flow and takes more clock cycles than other basic operations~\cite{10.1145/243439.243481}, which means it is unsuitable to have a direct hardware implementation in programmable switches.
For some use cases, division can be implemented by converting the dividend to its reciprocal at the end-host and then multiplying in the switch.

\niparagraph{Logarithms.}
The core operation of a floating point logarithm is the integer logarithm of the mantissa.
As prior research~\cite{5470752,201474,10.1145/98267.98294} shows, this can be done by a lookup table of fewer than 2000 entries with low error (<1\%). 

\niparagraph{Square roots.}
Square roots are even more expensive and time-consuming (\eg, more than 20 clock cycles) than division~\cite{10.1145/243439.243481,762835,624623}.
As with logarithms, we suggest a lookup-table-based approximation for this algorithm. 

\section{\arch Tofino resource utilization}
\label{sec:resource}

\tabref{tab:res} shows the resource utilization of \arch module out of a single Tofino pipeline. 
Most of these resources cannot be shared across multiple \arch instances. 
In particular, the high per-stage VLIW instruction utilization results from the need to implement variable-length shifts as multiple fixed-length shift operations, and is the limiting bottleneck that prevents multiple \arch instances from sharing a pipeline.

\begin{table}[!tb]
  \centering
  \caption{\arch resource utilization. Nine pipeline stages (out of 12 in total) are used.}
  \vspace{-1ex}
\scriptsize
  \label{tab:res}
  \begin{tabular}{m{2.7cm}rr}
    \toprule
   \bf Resource &   \bf Total usage & \bf Max usage in a MAU\\
    \midrule
  SRAM & 1.15\% & 5.00\%\\
  TCAM & 0.03\% & 4.17\%\\
  Stateful ALU 
& 8.33\% & 50.00\%\\
  VLIW instruction slots & 19.01\% & 96.88\% \\
  Input crossbar & 0.09\% & 4.38\%\\
  Result bus & 2.34\% & 12.50\% \\
  Hash bit & 1.06\% & 7.93\% \\
    \bottomrule
  \end{tabular}
  \vspace{-5ex}
\end{table}